\documentclass[a4paper,fleqn,usenatbib,useAMS]{mnras}
\usepackage{psfig}
\usepackage{hhline}
\usepackage{xspace}
\usepackage{natbib}
\usepackage{bigstrut}

\usepackage{graphicx}
\usepackage{amsmath}
\usepackage{xcolor}
\usepackage{bm}
\usepackage{times}
\usepackage{amssymb,amsmath}

\newcommand{\RXTE}{\textit{RXTE}\xspace}
\newcommand{\MAXI}{\textit{MAXI}\xspace}

\newcommand{\Swift}{\textit{Swift}\xspace}
\newcommand{\Swifts}{\textit{Swift's}\xspace}

\newcommand{\nH}{\ensuremath{N_\mathrm{H}}}
\newcommand{\ISIS}{\textsc{ISIS}\xspace}
\newcommand{\Chromos}{\textsc{Chromos}\xspace}
\newcommand{\HEASARC}{\textsc{heasarc}\xspace}
\newcommand{\HEASOFT}{\textsc{heasoft}\xspace}
\newcommand{\RN}[1]{%
  \textup{\uppercase\expandafter{\romannumeral#1}}%
}

\usepackage{times}
\usepackage{amssymb,amsmath}

\def\lsim{ \lower .75ex\hbox{$\sim$} \llap{\raise .27ex \hbox{$<$}} }
\def\gsim{ \lower .75ex \hbox{$\sim$} \llap{\raise .27ex \hbox{$>$}} }

\title[Modelling the evolving jet of MAXI J1836-194]{Correlating spectral and timing properties in the evolving jet of the micro blazar MAXI J1836-194}

\author[Lucchini et al.]
{M. Lucchini$^1$, T. D. Russell$^1$, S. B. Markoff$^{1,2}$, F. Vincentelli$^3$, D. Gardenier$^{1,4}$,\newauthor  C. Ceccobello$^5$, P. Uttley$^1$\\
$^1$API -- Anton Pannekoek Institute for Astronomy, University of Amsterdam, Science Park 904, 1098 XH Amsterdam, the Netherlands\\
$^2$GRAPPA -- Gravitational and Astroparticle Physics Amsterdam, University of Amsterdam, Science Park 904, 1098 XH Amsterdam, the Netherlands\\
$^3$Department of Physics and Astronomy, University of Southampton, SO17 1BJ, UK\\
$^4$ASTRON -- the Netherlands Institute for Radio Astronomy, Oude Hoogeveensedijk 4, 7991 PD, Dwingeloo, The Netherlands\\
$^5$Department of Space, Earth and Environment, Chalmers University of Technology, Onsala Space Observatory, 439 92 Onsala, Sweden
}

\begin{document}

\maketitle

\begin{abstract} 
During outbursts, the observational properties of black hole X-ray binaries (BHXBs) vary on timescales of days to months. These relatively short timescales make these systems ideal laboratories to probe the coupling between accreting material and outflowing jets as a the accretion rate varies. In particular, the origin of the hard X-ray emission is poorly understood and highly debated. This spectral component, which has a power-law shape, is due to Comptonisation of photons near the black hole, but it is unclear whether it originates in the accretion flow itself, or at the base of the jet, or possibly the interface region between them. In this paper we explore the disk-jet connection by modelling the multi-wavelength emission of MAXI J1836-194 during its 2011 outburst. We combine radio through X-ray spectra, X-ray timing information, and a robust joint-fitting method to better isolate the jet's physical properties. Our results demonstrate that the jet base can produce power-law hard X-ray emission in this system/outburst, provided that its base is fairly compact and that the temperatures of the emitting electrons are sub-relativistic. Because of energetic considerations, our model favours mildly pair-loaded jets carrying at least 20 pairs per proton. Finally, we find that the properties of the X-ray power spectrum are correlated with the jet properties, suggesting that an underlying physical process regulates both.
\end{abstract}



\maketitle

\section{Introduction}
Low mass black hole X-ray binaries are a class of binary systems in which a stellar-mass black hole is accreting mass from a low-mass companion star. Typically, these objects (BHXBs for brevity) are transient sources: they spend most of their lifetime in a faint quiescent state, occasionally shifting into outburst phases lasting a few weeks to a few months \citep{Tatarenko16}. When this happens, galactic BHXBs become some of the brightest sources in the X-ray sky. 

During a full outburst most BHXBs exhibit consistent behaviour, in the form of transitions between spectral states (see \citealt{Homan05,Remillard06} for reviews, and \citealt{Chen97} for a discussion of the variation among outbursts). As the source increases its luminosity while transitioning out of quiescence, its X-ray spectrum is dominated by a hard power-law component, originating in a yet poorly understood ``corona'' close to the black hole; this is defined as the hard state (HS from now on). "As the peak X-ray luminosity of the outburst is approached, the spectrum softens and becomes increasingly dominated by a black-body component associated with the accretion disk, while the power-law weakens or is completely absent. This is defined as the soft state, while transitional states are referred to as intermediate states (HIMS or SIMS respectively, depending on whether the source is closer to the hard or soft state). As the outburst decays and the luminosity decreases, the source eventually transitions back to the HS and then fades into quiescence. Furthermore, the properties of the X-ray light-curve are strongly correlated  with spectral properties \citep[e.g.,][]{Homan01,Homan05,Heil15b}, with harder states generally displaying more variability than softer ones. X-ray variability therefore provides an independent estimator of the properties of a source during an outburst \citep[e.g.,][]{Pottschmidt00,Pottschmidt03,Belloni05,Cassatella12,Heil15a}. Steady compact jets are detected in the HS and (typically) quenched in the soft state; as the source transitions from one to the other, individual knots of plasma are ejected from the system (e.g. \citealt{Fender04a}). A significant fraction ($\approx$ 40$\%$) of outbursts ``fail'' and only exhibit HSs \citep{Tatarenko16}.

Observationally, the properties of jet and accretion flow appear to be connected. In BHXBs, this takes the form of a correlation (in the HS) between the radio and X-ray luminosities \citep[e.g.,][]{Hannikainen98,Corbel00,Corbel03,Corbel13,Gallo03,Gallo14} which probe regions in the outer jet or near the black hole, respectively. The radio/X-ray correlation can be extended to accreting super-massive black holes in jetted active galactic nuclei (AGN) by including a term accounting for the black hole mass; this extension takes the form of a two-dimensional plane in the three-dimensional phase-space connecting radio luminosity, X-ray luminosity and black hole mass (\citealt{Merloni03}, \citealt{Falcke04}, \citealt{Plotkin12}). This plane is called the fundamental plane of black hole accretion; its existence implies that to first approximation black hole physics are scale-invariant \citep{Heinz03}, and what is learnt from one class of systems could potentially be applied to the other. However, the physics driving this connection between accreting and outflowing materials is still poorly understood. Thanks to their quick evolution, BHXBs are ideal laboratories to probe the disk-jet connection, and how it might be changing as a function of accretion rate.

Currently there is significant debate about the exact nature of the coronal power-law component. In general, coronal emission is believed to originate near the black hole due to inverse Comptonisation of soft disk photons by a population of hot electrons. Models of the corona can be broadly categorised in three groups,  invoking either a) the innermost regions of a radiatively inefficient accretion flow (RIAF, e.g. \citealt{Narayan94,Yuan07}) b) a slab corona extending over the accretion disk \citep[e.g.,][]{Haardt93,Haardt94} or c) a compact location above the black hole (the so-called lamp-post geometry, e.g. \citealt{Matt91,Matt92,Martocchia96,Beloborodov99}) as the location of the Comptonised emission. In the latter scenario, the base of a jet is often invoked as a natural physical realisation of the lamp-post (e.g. \citealt{Markoff05,Maitra09,Dauser13,Kara19}). Additional contributions from non-thermal synchrotron emission originated in the jet have also been invoked (e.g.  \citealt{Markoff01}).  

In jet models, the emitting leptons are typically assumed to be fully relativistic throughout the outflow, starting from the base (e.g. \citealt{Falcke95}, \citealt{Markoff01}, \citealt{Potter12}, \citealt{Malzac14}). \cite{Connors19} highlighted that this assumption has a large impact on the Comptonisation spectra from their launching regions (or in other words, from the ``lamp-post like'' region), and is in tension with observations if one is to fit the full X-ray spectrum without additional contributions, like non-thermal synchrotron emission from the jet. If the electron distribution is relativistic (its temperature is $T_{\rm e}\gg 511
\rm{keV}$ or equivalently the minimum lepton Lorenz factor is $\gamma_{\rm min}\gg 1$), then fitting the data requires the X-ray emitting regions to be extended and very optically thin ($\tau \leq 0.01$). In this regime, the Comptonisation spectrum is produced in a single scattering, rather than a superposition of many scatterings. The resulting X-ray spectrum shows significant curvature, and is unlike a typical power-law continuum produced by many consecutive scatterings in more optically thick ($\tau \approx 0.1-1$) media \citep[e.g.,][]{Grebenev93,Zdziarski97,Barret00,Joinet07}. The only way to produce a power-law X-ray emission with these parameters is to fine tune different spectral components, possibly originating in different regions of the system. As a result, past works often invoked a mix of synchrotron and Comptonisation from two different emitting regions in order to match the X-ray spectra of BHXBs (e.g. \citealt{Markoff05,Nowak11}), particularly the hardening above 10 keV: the hard X-rays are due to Comptonisation in the jet base, while the soft X-rays are due to synchrotron emission produced downstream in the jet. This combination also resulted in lower reflection fractions as some of the hardening was absorbed into the continuum. \cite{Connors19} pointed out that this is inconsistent with the observed hard lags in XRBs (\citealt{Kotov01}, \cite{Arevalo06}). If instead the synchrotron emission does not extend all the way to the soft X-rays, \cite{Connors19} showed that the base of the jet can contribute a few to to $\approx 50\%$ of the X-ray flux, but the bulk of it has to be produced in a different region (such as the hot accretion flow). In this paper we show that this need not be the case if the electrons in the jet base are in the mildly-relativistic regime, in which $T_{\rm e}\leq 511\,\rm{keV}$, and contains a moderate ($\approx$ 20) number of electron-positron pairs for each proton.

\begin{table}
\begin{center}
\begin{tabular}{lcc}
\hline
Instrument & ObsID & Date and time\\
\hline
\RXTE/PCA & 96371-03-01-00 & \textbf{Aug 31, 11:09 - 15:54}\\
\RXTE/PCA & 96371-03-02-00 & Sep 01, 13:28 - 16:06\\
\RXTE/PCA & 96371-03-03-00 & Sep 02, 11:30 - 13:21\\
\RXTE/PCA & 96371-03-03-01 & Sep 04, 07:23 - 08:41\\
\Swift/XRT & 00032085001 & \textbf{Aug 30, 16:47 - 17:04}\\
\hline
\RXTE/PCA & 96438-01-01-04 & \textbf{Sep 14, 20:12 - 20:31}\\
\Swift/XRT & 00032087009 & \textbf{Sep 13, 06:54 - 21:34}\\
\hline
\RXTE/PCA & 96438-01-02-03 & \textbf{Sep16, 09:13: - 10:25}\\
\RXTE/PCA & 96438-01-02-00 & Sep 17, 20:13 - 20:51\\ 
\Swift/XRT & 00032087012 & \textbf{Sep 16, 02:05 - 02:28}\\
\hline 
\RXTE/PCA & 96438-01-03-01 & Sep 25, 19:05 - 19:42\\
\RXTE/PCA & 96438-01-03-05 & \textbf{Sep 26, 21:21 - 21:34}\\
\RXTE/PCA & 96438-01-03-02 & Sep 27, 19:51 - 20:55\\
\Swift/XRT & 00032087017 & \textbf{Sep 26, 04:14 - 19:16}\\
\hline
\RXTE/PCA & 96438-01-05-02 & Oct 11, 18:33 - 18:57\\
\RXTE/PCA & 96438-01-05-05 & \textbf{Oct 12, 13:21 - 13:55}\\
\RXTE/PCA & 96438-01-05-06 & Oct 13, 19:06 - 19:26\\
\Swift/XRT & 00032087024 & \textbf{Oct 12, 20:06 - 20:23}\\
\hline
\RXTE/PCA & 96438-01-12-01 & Oct 26, 06:18 - 07:18 \\
\RXTE/PCA & 96438-01-12-02 & \textbf{Oct 27, 17:45 - 18:14} \\ 
\RXTE/PCA & 96438-01-12-03 & Oct 28, 09:51 - 10:10 \\
\Swift/XRT & 00032087024 &  \textbf{Oct 27, 05:06 - 11:45}\\
\hline
\end{tabular}
\caption{X-ray data used in this work. The epochs considered in the spectral analysis have their dates shown in bold font; all the others were only used in the timing analysis.}
\label{tab:rxte}
\end{center}
\end{table}

\begin{figure*}
\hspace{-0.5cm}
\includegraphics[width=1.0\textwidth]{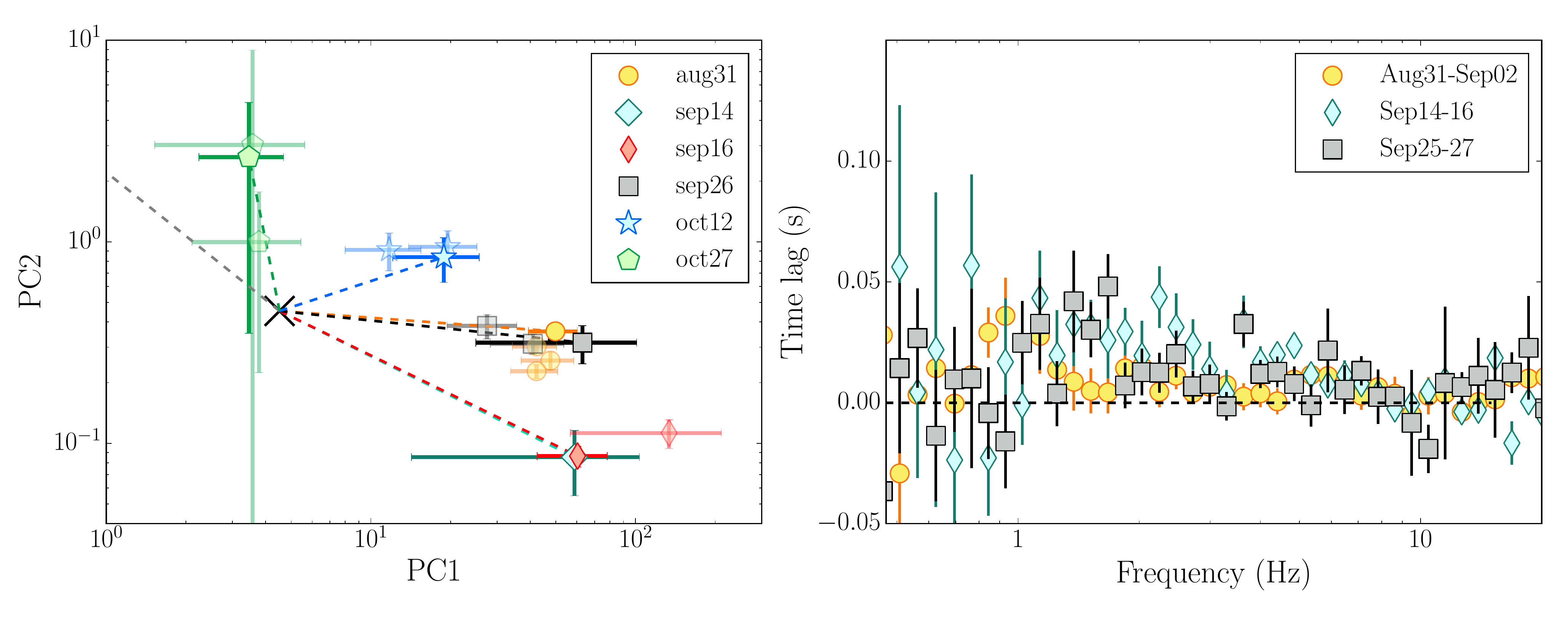}
\caption{Evolution of X-ray timing properties of the source during the outburst. The left panel shows the evolution of the source power spectra in the (power) colour-colour diagram \citep{Heil15a}. Opaque points indicate the \RXTE observation closest to \Swift coverage; transparent ones indicate all the other epochs reported in Tab.\ref{tab:rxte}. The right panel shows our tentative ($\approx 2-3\sigma$) detection of a hard lag between the 3.5-5 and 7-13 keV bands, with hard photons lagging soft ones by $\approx 10^{-2}\,\rm{s}$.} 
\label{fig:timing}
\end{figure*}
\begin{figure}
\hspace{-0.5cm}
\includegraphics[width=0.48\textwidth]{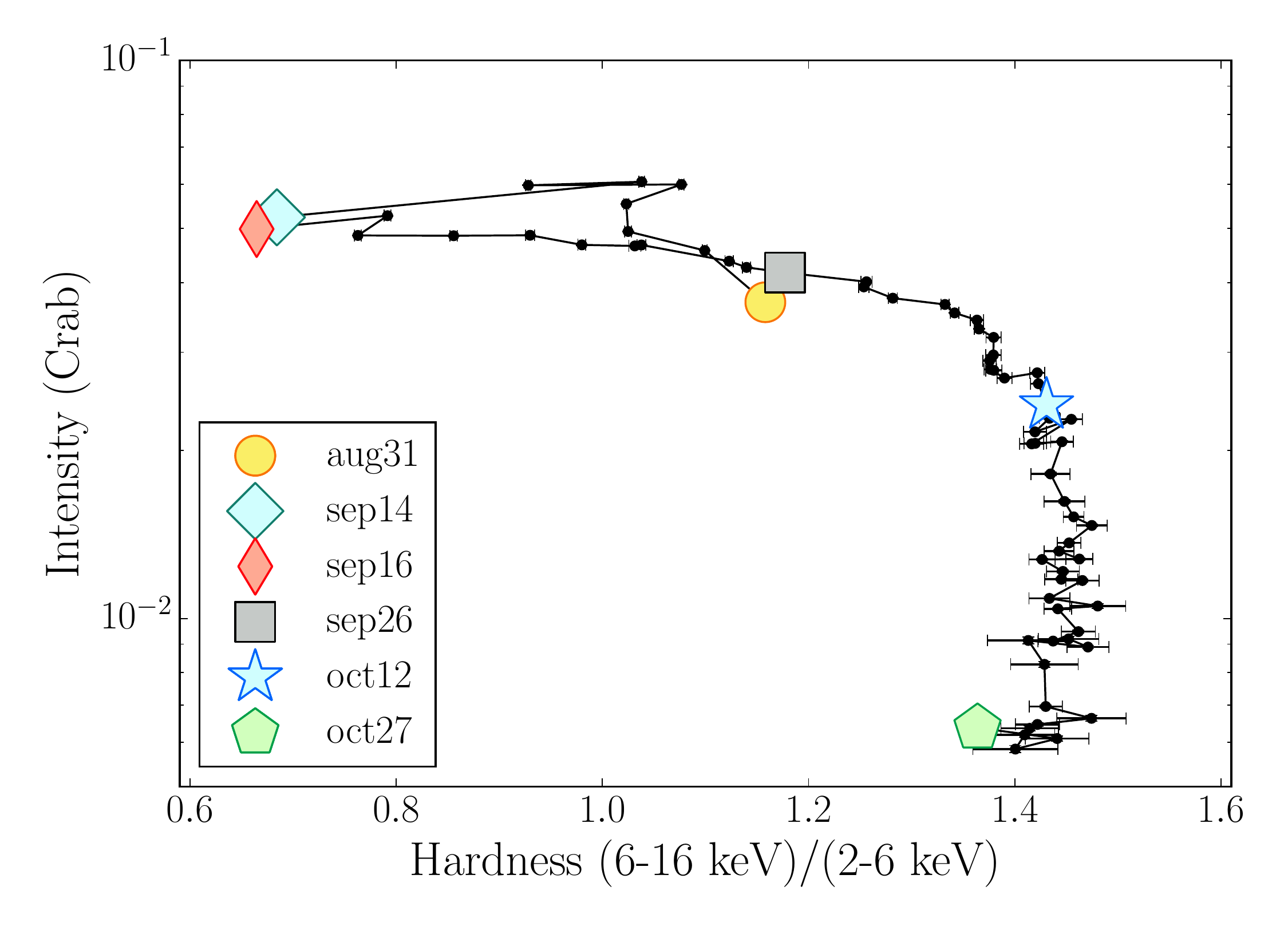}
\caption{Hardness/intensity diagram of \MAXI J1836$-$194 during its outburst; hardness is defined as the ratio between 6-16 and 2-6 keV count rates.} 
\label{fig:hid}
\end{figure}

MAXI J1836$-$194 is a low mass BHXB that went in outburst during late August 2011; it was quickly identified as a black hole candidate (\citealt{Negoro11}, \citealt{Strohmayer11}, \citealt{Miller-Jones11}, \citealt{Russell11}). Later studies of the source's X-ray spectra found that the black hole spin is likely high ($\approx 0.9$), and that the hard X-rays show a very prominent reflection component (\citealt{Reis12}, \citealt{Dong20}). Rather than undergoing a full outburst, the source reached a hard-intermediate state (HIMS) before going back into the HS and fading into quiescence \citep{Ferrigno12}. The source was the target of an extensive multi-wavelength campaign, which produced some of the best spectral coverage of a BHXB outburst to date (\citealt{Russell13}, \citealt{Russell14b}). \cite{Peault19} modelled its evolving jet by combining the radio-through-infrared SEDs with the X-ray power spectra, finding that the jet emission is consistent with an internal shock scenario. \MAXI J1836$-$194 is unique in that its viewing angle ($4^{\circ}<\theta<15^{\circ}$, \citealt{Russell14a}) is the lowest in the X-ray binary population to date, and is thus comparable to many jet-dominated AGN such as M87 or (some) blazars. The low viewing angle, combined with the excellent multi-wavelength data, makes it an excellent source to study both the disk-corona-jet connection and scale-invariant models of black hole accretion. 

The goal of this paper is to quantify the conditions required for the base of the jet to comprise the X-ray emitting corona, particularly in terms of size, particle temperature and mass content. We do so by modelling the six multi-wavelength SEDs presented in \cite{Russell14b} with the \texttt{bljet} disk+jet model (\citealt{Lucchini19a}, Paper \RN{1} from now on), which in itself is an extension of the \texttt{agnjet} code (\citealt{Markoff01}, \citealt{Markoff05}, \citealt{Maitra09}, \citealt{Connors19}). \texttt{bljet} was originally developed  for modelling AGN SEDs, and is applied to BHXBs here for the first time. Furthermore, we improve the broadband coverage of the source by analysing additional hard X-ray spectral and timing data gathered by \RXTE. Finally, we combined timing and spectral information in our modelling with the goal of providing a clearer picture of the coupling between these two observables. The paper is structured as follows. In section \ref{sec:data} we describe the analysis of the data used, in section \ref{sec:model} we discuss our full disk+jet multiwavelength model and apply it to the data, in section \ref{sec:discussion} we discuss our findings, and in section \ref{sec:conclusion} we draw our conclusions.

\section{Data analysis}
\label{sec:data}

In this paper we model the six quasi-simultaneous SEDs from \cite{Russell14b}, including additional hard X-ray data provided by \RXTE. The details of the radio and \Swift/XRT data reduction are presented in \cite{Russell14b}. The infrared, optical and UV data were first presented in \cite{Russell13}. We note that in this paper we label epochs based on the date of the \RXTE/PCA pointings, while \cite{Russell14b} used the date of the radio observations instead.

\subsection{RXTE data reduction}

We searched the \HEASARC archives for \RXTE observations close to our \Swift/XRT observations. The resulting ObsIDs are reported in Tab.\ref{tab:rxte}. We used the standard \RXTE tools in \HEASOFT, version 6.26.1, through the \Chromos\footnote{https://github.com/davidgardenier/chromos} pipeline (\citealt{Gardenier18}, which also contains the details of the data reduction process) to extract PCA energy spectra and light-curves. The spectra were extracted from standard-2 events in the proportional counter unit 2 (PCU2) only. The light-curves were extracted from good-xenon events in the $3-13$, $3-5.5$ and $7-13$ keV bands. We jointly modelled the \Swift/XRT and only the \RXTE/PCA spectra closest in time to each \Swift observation, but analysed the light-curves of every observation.

\begin{figure}
\includegraphics[width=0.45\textwidth]{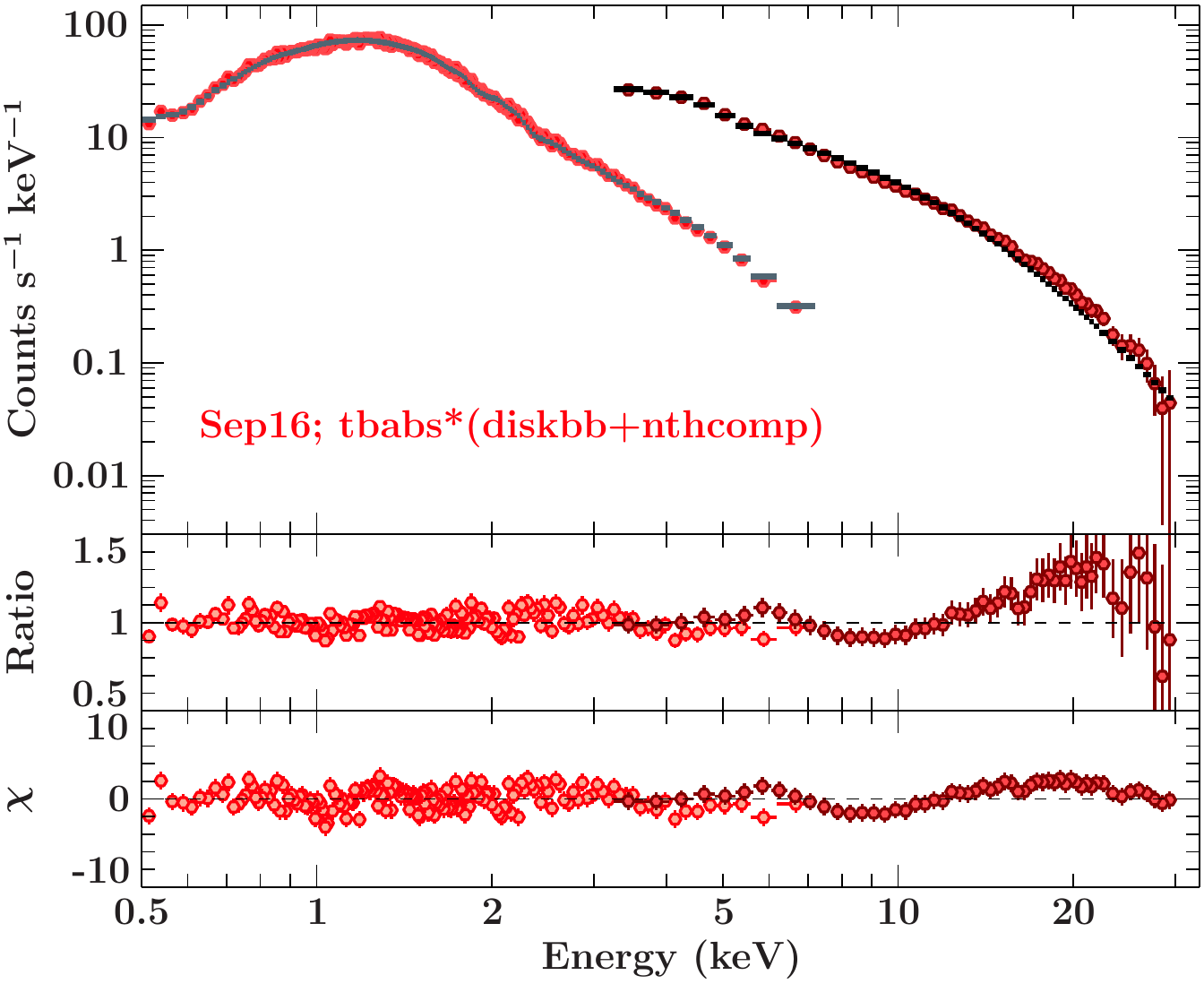}
\caption{Sep16 \Swift/XRT (light red) and \RXTE/PCA (dark red) spectra fitted without a disk reflection contribution. As shown by the residual and ratio plots, the characteristic reflection signatures are very prominent in the \RXTE band.}
\label{fig:sep16}
\end{figure}

\subsection{Timing analysis}

We investigated the X-ray variability of the source by analysing its \RXTE lightcurves using Fourier domain techniques. The \Chromos pipeline automatically computes the noise-subtracted averaged power spectrum (PSD), correcting for background and dead time detector effects, as well as the power colours of the PSD for each observation chosen. Power colours are the ratio of the integral of the PSD in four bands (band 1: $0.0039-0.031$ Hz, band 2: $0.031-0.25$ Hz, band 3: $0.25-2.0$ Hz, band 4: $2.0-16.0$ Hz); the two power colours are defined as band 3/band 1 and band 2/band 4, respectively, and can be thought of as a way to quantify the shape of the PSD regardless of its normalisation. This calculation is identical to that presented in \cite{Gardenier18}. The evolution of the source in the power colour-colour diagram is shown in the left panel of Fig. \ref{fig:timing}. The power spectral hue, as in \cite{Heil15a}, is defined as the angle between the location of each point in the power colour-colour diagram, and a line angled $45^{\circ}$ toward the top left of the panel, shown in Fig.\ref{fig:timing} by the grey dashed line. The spectral evolution of the source, presented in \cite{Russell15}, in the hardness-intensity diagram is shown in Fig.\ref{fig:hid}. The timing and spectral properties of the source are well correlated, with epochs with larger hues being brighter and softer.

\begin{figure*}
\caption{Ratio (data/model) and residual (data-model/error) plots for the X-ray spectra of the source throughout the outburst. On August 31st, October 12th and October 27th the model is in good agreement with the data. On September 13 and 16, the \Swift/XRT spectra appear to require a softer power-law than the \RXTE/PCA ones, resulting in residuals above 10 keV. On September 16 and 26 there are systematic residuals between 4 and 7 keV, likely caused by the simple reflection model used here. The $\chi^{2}/{\rm dof}$ are: 52.63/55, 337.98/257, 328.01/224, 121.28/115, 68.68/74, and 44.04/63, respectively.} 
\includegraphics[width=0.48\textwidth]{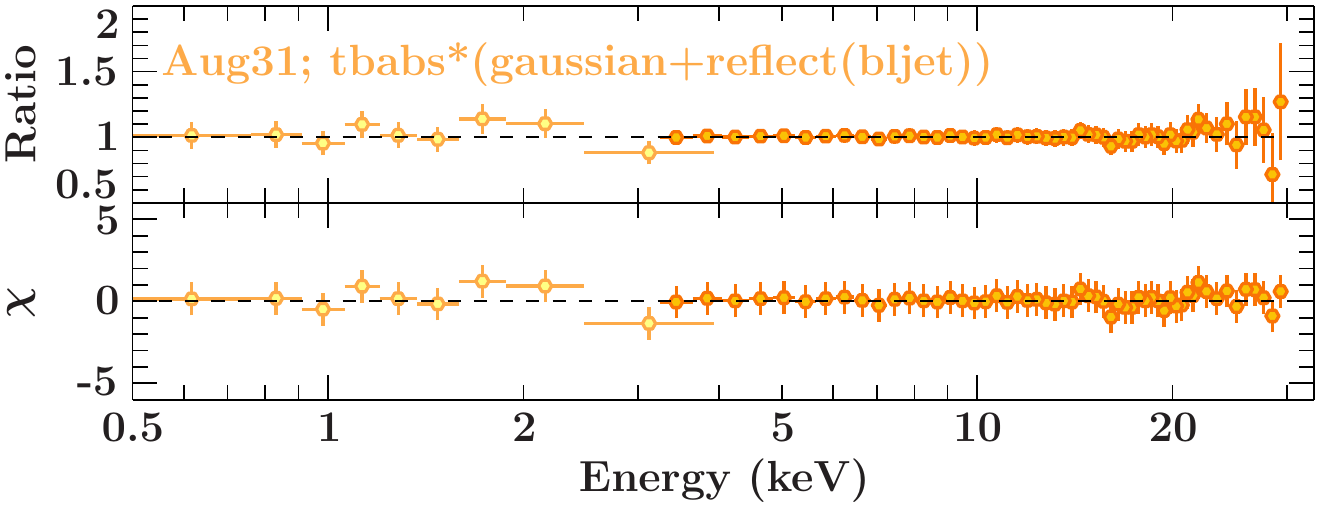}\quad
\includegraphics[width=0.48\textwidth]{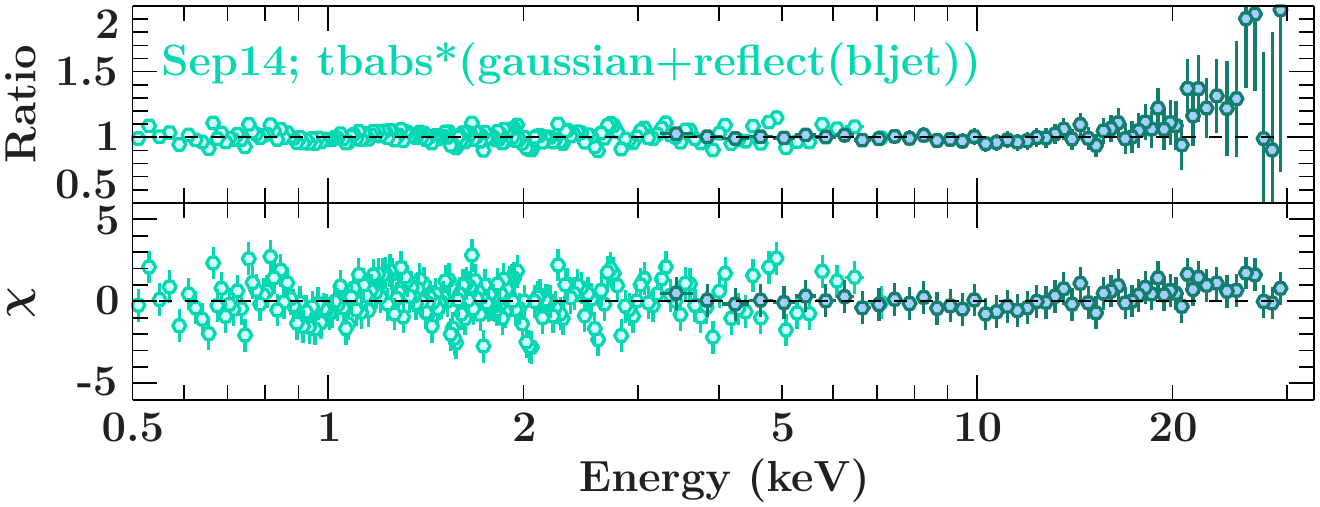}
\includegraphics[width=0.48\textwidth]{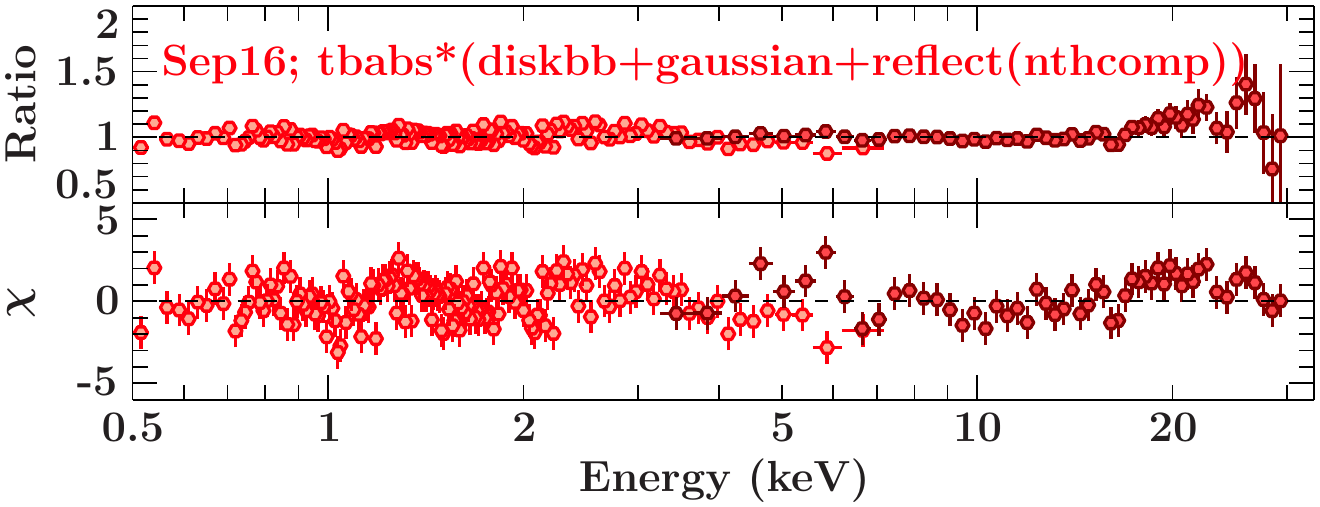}\quad
\includegraphics[width=0.48\textwidth]{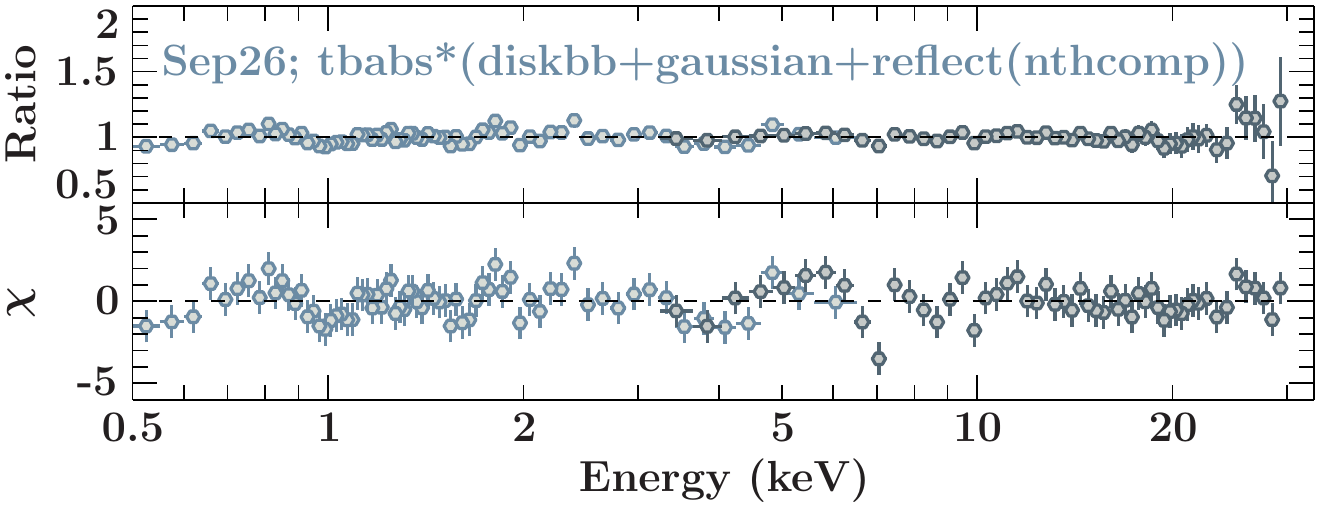}
\includegraphics[width=0.48\textwidth]{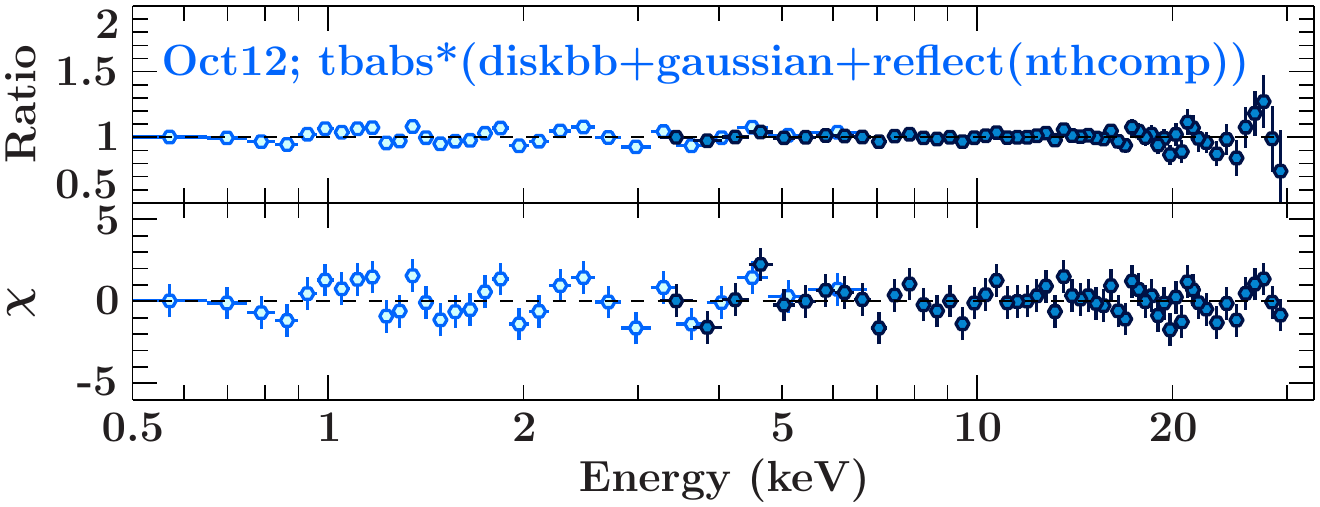}\quad
\includegraphics[width=0.48\textwidth]{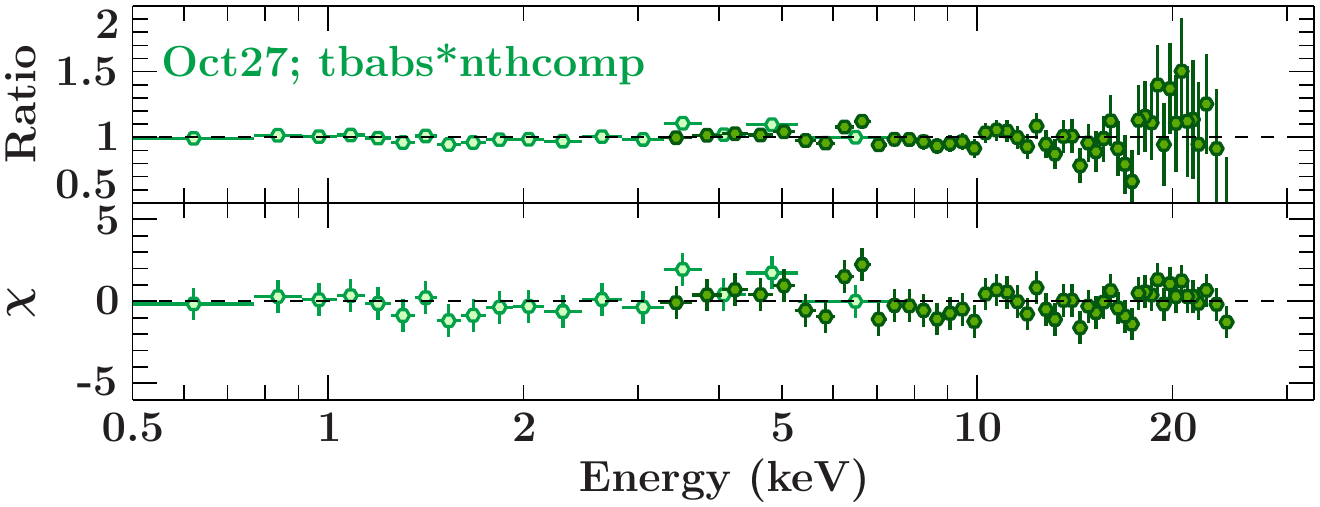}
\label{fig:residuals}
\end{figure*}
\begin{table*}
\begin{tabular}{cccccccccc}
Epoch & Constant & \nH & \texttt{Diskbb} & $kT_{\rm{disk}}$ & \texttt{nthcomp} & $\Gamma$ & relfrac & \texttt{Gaussian} & \texttt{Gaussian}  \\ 
 &  & $10^{22}$ cm$^{-2}$ & Norm ($10^{3}$) & (keV) & Norm &  &  & Norm ($10^{-4}$) & $\sigma$ (keV) \\ \hline
Aug 30 & $1.15^{+0.07}_{-0.09}$ & $0.38^{+0.09}_{-0.07}$ & $62^{+73}_{-44}$ & $0.19^{+0.03}_{-0.02}$ & $0.27^{+0.02}_{-0.03}$ & $2.10^{+0.01}_{-0.02}$ & $0.80^{+0.11}_{-0.07}$ & $7.2^{+1.8}_{-1.3}$ & $0.78^{+0.24}_{-0.19}$ \\
Sep 13 & $0.90^{+0.01}_{-0.01}$ & $0.36^{+0.01}_{-0.01}$ & $5.64^{+0.31}_{-0.27}$ & $0.40^{+0.01}_{-0.01}$ & $0.77^{+0.02}_{-0.03}$ & $2.85^{+0.04}_{-0.05}$ & $2.74^{+0.40}_{-0.36}$ & $1.8^{+2.4}_{-1.3}$ & $1.4^{+2.6}_{-0.9}$ \\
Sep 16 & $0.94^{+0.01}_{-0.01}$ & $0.396^{+0.005}_{-0.005}$ & $8.43^{+0.38}_{-0.35}$ & $0.414^{+0.004}_{-0.004}$ & $0.27^{+0.01}_{-0.01}$ & $2.88^{+0.03}_{-0.03}$ & $3.19^{+0.34}_{-0.30}$ & $2.1^{+2.4}_{-1.5}$ & $1.35^{+1.01}_{-0.49}$ \\ 
Sep 26 & $1.26^{+0.02}_{-0.02}$ & $0.36^{+0.02}_{-0.02}$ & $11.5^{+4.4}_{-2.7}$ & $0.25^{+0.02}_{-0.01}$ & $0.32^{+0.02}_{-0.02}$ & $2.34^{+0.07}_{-0.06}$ & $2.46^{+0.54}_{-0.52}$ & $36.9^{+5.9}_{-6.2}$ & $3.1^{+0.3}_{-0.4}$ \\ 
Oct 12 & $1.05^{+0.02}_{-0.02}$ & $ 0.35^{+0.03}_{-0.06}$ & $9.5^{+6.7}_{-5.6}$ & $0.18^{+0.02}_{-0.01}$ & $0.12^{+0.01}_{-0.02}$ & $1.81^{+0.04}_{-0.03}$ & $ 0.48^{+0.18}_{-0.11}$ & $7.4^{+5.6}_{-2.9}$ & $1.6^{+0.7}_{-0.5}$   \\
Oct 27 & $0.56^{+0.02}_{-0.01}$ & $0.31^{+0.02}_{-0.01}$ & // & // & $0.050^{+0.002}_{-0.001}$ & $1.69^{+0.02}_{-0.02}$ & // & // & // \\ \hline
\end{tabular}
\caption{Best fitting parameters for our phenomenological fits of the \Swift+\RXTE spectra.}
\label{tab:xfits}
\end{table*}

We also computed the Fourier time lag of the 7-13 keV lightcurve with respect to the 3-5.5 keV lightcurve following the recipe described in \cite{Uttley14}, through the timing software software \textsc{stingray} \citep{stingray}. In  particular we used 8192 bin per segments and logarithmic binning factor of 1.1. In order to increase the statistics we stacked the lightcurves in three groups: one with all the observations between August 31 and September 2, a second one with data collected between September 14 and 16, and a last one considering the all observations between September 25 and 27. We note that these epochs have power colours consistent with each other, indicating that the variability properties of the source are relatively similar in the three periods chosen. Results are shown in the right panel of Fig. \ref{fig:timing}. Even though the statistics are poor, we found evidence ($\approx 2-3\sigma$) for a hard lag between 1 and 10 Hz for all of the early part of the outburst. In the latter epochs, the statistics are insufficient to detect such a hard lag.

This hard lag is consistent with the standard propagating fluctuations model (e.g. \citealt{Lyubarskii97}), and is a first hint that the X-ray emitting region is located close to the black hole, rather than downstream ($\approx 10^{3}\,\rm{R_g}$) in the jet.

\subsection{X-ray spectral fits}
\label{sec:Xrayfits}

All spectral fits in this work are performed in \ISIS version 1.6.2-35 \citep{Houck00}. The data are fit first by running the \texttt{subplex} least-$\chi$ squared algorithm to get close to a good fit, and then by running the \texttt{emcee} Markov Chain Monte Carlo algorithm \citep{emcee}. We use 20 walkers per free parameter for each chain. All the walkers are initialised uniformly within 1\% of the best-fit values found by \texttt{subplex}. We define the chain as having converged after it reaches the point when a) the acceptance rate of the chain stabilises and b) the posterior distribution is unchanged. This results in chain lengths of typically a few thousand; depending on the behaviour of the posterior distribution in each fit, taking the initial burn-in period to be between 30-75\% of the chain. We define the value of the best-fitting parameter as the median of the walker distribution after the chain has converged, as we found that this produces a better fit to the data than the peak of the posterior. We define the 1-$\sigma$ uncertainty as the interval in the posterior distribution which contains 68\% of the walkers, after excluding the burn-in period.

\Swift/XRT spectra were re-binned to a minimum signal-to-noise ratio of 20 per bin in every epoch except on August 31st. In this epoch, due to the lack of statistics, we only used a minimum signal-to-noise ratio of 10. In every \Swift/XRT spectrum we kept data between 0.5 and 10 keV. \RXTE/PCA spectra were re-binned to a minimum signal-to-noise ratio of 4.5 per bin in order to use $\chi^{2}$ statistics. Depending on the quality of the data, we either considered data between 3 and 20 keV (on October 27th) or 3 and 30 keV (every other epoch). In every epoch we added a $1\%$ systematic uncertainty to the \RXTE/PCA data in order to account for cross-calibration uncertainties. Furthermore, when modelling the \RXTE spectra we multiplied the data by a constant in order to account for the observations not being strictly simultaneous, as well as any additional cross-calibration uncertainties.

Before applying our physical multi-wavelength model to the entire data set, we fit the X-ray spectra alone with phenomenological models in order to gain a better understanding of the broad properties of the system. In particular, we aim to constrain the origin of the coronal X-ray  emission. In the framework of a jet-dominated model, two radiative mechanisms are viable candidates to produce X-ray emission: inverse Compton scattering of disk and cyclo-synchrotron photons in the jet base, located close to the black hole, or non-thermal synchrotron emission originating further downstream in the jet. In the former, much stronger reflection features are predicted than in the latter \citep{Markoff04}; while typical sources favour the inverse Compton+reflection scenario (e.g. GX\,339$-$4; \citealt{Connors19}), exceptions do exist, such as XTE\,1118$+$480 \citep{Miller02,Maitra09} and XTE\,1550-564 \citep{Russell10}. To first order, one would expect that a low inclination source like MAXI\,J1836$-$194 would be a prime candidate for detection of the non-thermal synchrotron component, as it originates in regions further out in the jets, that are more beamed than the jet base.

We first fit the spectra with an (absorbed) comptonisation continuum (\texttt{nthcomp}; \citealt{nthcomp}) and a standard accretion disk component (\texttt{diskbb}; \citealt{diskbb}), with the exception of the October 27th data for which we only use \texttt{nthcomp} (no disk contribution was found in the soft X-rays). In order to account for the non-simultaneity of the \Swifts/XRT and \RXTE/PCA observations we included a multiplicative calibration constant between the data sets; the final syntax of the model is \texttt{constant*tbabs*(nthcomp+diskbb)} or \texttt{constant*tbabs*(nthcomp)}, respectively. We freeze the constant to 1 for the \Swifts/XRT data and leave it free for the \RXTE/PCA data. This model fits the soft X-ray data fairly well, but significant residuals are present around 6 and above 10 keV in every epoch except October 27th; Fig.\ref{fig:sep16} shows this for the data on September 16th. Both of these are clear signs of reflection, indicating that the source of the X-rays is located close to the black hole and accretion disk. 

We model these signatures with a Gaussian centred at 6.4 keV and the \texttt{reflect} model of \cite{reflect}. The syntax for the final X-ray spectral model in epochs that show reflection signatures is \texttt{constant*tbabs*(reflect(nthcomp)+diskbb+\\gaussian)}. This model provides a fair description of the data; the residuals for each epoch are shown in Fig.\ref{fig:residuals} and the best-fitting parameters along with uncertainties are reported in Tab.\ref{tab:xfits}. However, on the spectra of September 13, 16 and 26 residual features remain above 10 keV, and between 4 and 7 keV. We stress that the goal of these phenomenological spectral fits is simply to constrain the location of the X-ray emitting region by quantifying the presence or absences of reflection features. These phenomenological spectral fits favour the inverse Compton scenario, as the reflection features are very strong, particularly near the peak of the outburst, in agreement with other works (\citealt{Reis12}, \citealt{Dong20}). This picture is further strengthened by our tentative detection of a hard lag in the X-ray lightcurves.

\section{Multi-wavelength model}
\label{sec:model}

Next, we model the six multi-wavelength SEDs with an updated version of the multi-zone jet model \texttt{bljet}, the full details of the original model are presented in Paper \RN{1}. Briefly, \texttt{bljet} originated as an extension of the \texttt{agnjet} model of \cite{Markoff05}; it is designed to mimic the conservation of energy within the jets similar to ideal MHD, and now confirmed by GRMHD simulations. In this scenario, the jet is launched with a high initial magnetisation, and accelerates by converting its Poynting flux into bulk kinetic energy \citep[e.g.,][]{McKinney06,Chatterjee19}. Unlike \texttt{agnjet}, which is limited to mildly relativistic outflows, the jet in \texttt{bljet} can reach arbitrarily high Lorenz factors. 

The model assumes that a fraction $N_{\rm j}$ of the Eddington luminosity is injected at the jet base near the black hole, the resulting internal energy is divided between magnetic fields, cold protons and thermal leptons. This injection occurs in a cylindrical region characterised by an initial radius $r_{\rm 0}$, with an aspect ratio $h = R_{\rm 0}/z_{\rm 0} = 2$ as in papers 1 and 2. The jet then accelerates up to a final speed $\gamma_{\rm f}$ by converting its initial magnetic field into bulk kinetic energy; this occurs at a distance $z_{\rm acc}$, where the magnetisation reaches a value $\sigma_{\rm f} \leq 1$. The jet is assumed to have a parabolic shape in the bulk acceleration region, and a conical shape afterwards, in agreement with VLBI observations of AGN \citep[e.g.,][]{Mertens16,Hada16}. The details of the magnetic-to-kinetic energy conversion and of the rest of the jet dynamics are detailed in Paper \RN{1}.

At a distance $z_{\rm diss}$ from the black hole, the jet experiences a dissipation region in which particle acceleration begins. From this point onward, a fraction (which initially we take to be $10\%$, as in Paper \RN{1} and \citealt{Lucchini19b}) of the leptons is channelled in a non-thermal tail. Following paper 1, we set $z_{\rm diss} = z_{\rm acc}$ in order to reduce the number of free parameters, but note that it does not always need to be the case \citep{Lucchini19b}.  

\subsection{Particle distributions}
\label{sec:particles}

Previous versions of \texttt{agnjet} and \texttt{bljet} assumed the particle distributions were always relativistic and thus calculated in energy space, meaning that the leptons are assumed to have an energy $E_{\rm e}$ and a corresponding Lorenz factor:
\begin{equation}
  \gamma_{\rm e} = E_{\rm e}/m_{\rm e}c^{2};  
  \label{elec_energy}
\end{equation}
the lepton particle distribution was defined as $N(E_{\rm e})$. This approximation is appropriate in the relativistic regime ($\gamma_{\rm e} \gg m_{\rm e}c^{2}$, implying $T_{\rm e} \gg 511\,\rm{keV}$), but it is incorrect if the electron temperature is non-relativistic, thus in older versions the minimum temperature allowed was $\sim 511$ keV. 

In this work, we have updated the calculations to be in momentum space in order to treat both relativistic and non-relativistic particle distributions. Similar to \cite{Ghisellini98}, we describe the leptons starting from their relativistic momentum $p_{\rm e}$, which corresponds to a Lorenz factor:
\begin{equation}
\gamma_{\rm e}(p) = \sqrt{\varrho^{2}+1},
\label{elec_p}
\end{equation}
where $\varrho = p/m_{\rm e}c$ is the electron momentum in units of $m_{\rm e}c$. Unlike equation \ref{elec_energy}, this expression is valid in both the relativistic ($p_{\rm e} \gg m_{\rm e}c^{2}$, or $\gamma_{\rm e}(p) \gg 1$) and non-relativistic ($p_{\rm e} \leq m_{\rm e}c^{2}$, or $\gamma_{\rm e}(p) \approx 1$) regimes. For the rest of this section, we will drop the subscript $e$, implicitly assuming that $\gamma$ and $\varrho$, as well as the temperature $T_{\rm e}$ and number density $n_{\rm e}$, refer to leptons.
 
As in older versions of the model, we assume that up to $z_{\rm diss}$ within the jet base, the particle distribution remains thermal and that the jet is isothermal, meaning that adiabatic losses are neglected and therefore $T$ is unchanged. In this inner jet region, the particle distribution is described by the Maxwell-J{\"u}ttner distribution:
\begin{equation}
    N(\gamma) = N_{\rm th,0} \varrho^2e^{-\frac{\gamma(\varrho)}{\theta}},
    \label{eq:MJ}
\end{equation}
where $N_{\rm th,0} = n/m_{\rm e}c^{3}\theta K_{\rm 2}(1/\theta)$ is the normalisation of the Maxwell-J{\"u}ttner so that its integral over particle momenta is $n$, and $\theta = k T/m_{\rm e}c^{2}$ the dimensionless temperature.

Beyond $z_{\rm diss}$, where we assume non-thermal particle acceleration begins, we calculate the radiating particle distribution by solving the steady state continuity equation along the length of the jet: 
\begin{equation}
    N(\varrho)= \frac{\int_{\varrho}^{\infty} Q(\varrho) d\varrho}{\dot{\varrho}_{\rm ad}+\dot{\varrho}_{\rm rad}},
\end{equation}
where $Q(\varrho)$ is the injection term, $\dot{\gamma}_{\rm rad}$ is the radiative loss term and $\dot{\gamma}_{\rm ad}$ is the adiabatic loss term. The injection term $Q(\varrho)$ is assumed to be a mixed distribution, composed by a Maxwell-J{\"u}ttner thermal distribution plus an exponentially-cutoff power-law tail:
\begin{equation}
Q(\varrho) = \left\{\begin{aligned} \frac{(1-f_{\rm pl})N_{\rm th,0}}{t_{\rm inj}}\varrho^2e^{-\frac{\gamma(p)}{\theta}}, & \quad \varrho<\varrho{\rm min, pl} \\
\frac{(1-f_{\rm pl})N_{\rm th,0}}{t_{\rm inj}}\varrho^2e^{-\frac{\gamma(p)}{\theta}} + & \\ \frac{f_{\rm pl}N_{\rm nth,0}}{t_{\rm inj}}\varrho^{-s}e^{-\frac{\gamma(p)}{\gamma(p)_{\rm max, pl}}},&\quad \varrho\geq \varrho_{\rm min, pl}  \end{aligned}\right.
\end{equation}\\
where $f_{\rm pl}$ is the fraction of particles channelled into the power-law, $s$ is the slope of the injected power-law, $N_{\rm th,0}$ is the normalisation of the thermal particles as in eq.\ref{eq:MJ}, $N_{\rm th,0} = (1-s)/(\varrho_{\rm max}^{1-s} - \varrho_{\rm min}^{1-s})$ is the normalisation of the power-law tail. The two distributions are normalised such that the total lepton number density $n$ is always $n = n_{\rm th} + n_{\rm pl}$. The injection time is defined as $t_{\rm inj} = r(z)/c$, where $r(z)$ is the radius of the jet and $c$ the speed of light. $\varrho{\rm min, pl}=\langle\varrho_{\rm th}(\theta)\rangle$ is the minimum of the non-thermal particle distribution, which is always assumed to be equal to the average dimensionless momentum of the thermal distribution for a given dimensionless temperature $\theta$. $\varrho{\rm max, pl}$ is the maximum particle momentum. The corresponding Lorenz factor from which it is calculated is derived identically to paper 1; for completeness this derivation is reported below. From each distribution in $\varrho$ space, the corresponding distribution in Lorenz factor space is always calculated as:
\begin{equation}
N(\gamma) = N(\varrho)\frac{d\varrho}{d\gamma}.    
\end{equation} 
The radiative loss term is defined as in \cite{Ghisellini98}:
\begin{equation}
    \dot{\varrho}_{\rm rad} = \frac{4 \sigma_{\rm t}c U_{\rm rad}}{3m_{\rm e} c^{2}}\varrho \gamma,
\label{eq:rad_dot}
\end{equation}
where $\sigma_{\rm t}$ is the Thomson cross section. In this paper, the radiative energy density $U_{\rm rad}$ includes only the synchrotron term, such that  $U_{\rm rad} = U_{\rm b}$. This is because at the distances over which particle acceleration occurs ($\approx 10^{3}-10^{6}\,R_{\rm g}$) radiative cooling is dominated by cyclo-synchrotron, and the disk contribution is negligible. 

Similarly, the adiabatic loss term is defined as:
\begin{equation}
    \dot{\varrho}_{\rm ad} = \frac{\beta_{\rm eff}c}{r}\varrho,
\label{eq:ad_dot}
\end{equation}
where $\beta_{\rm eff}$ is the effective expansion speed of the jet. If the jet is purely isothermal then adiabatic losses are assumed to be entirely balanced by some acceleration mechanism, and thus $\dot{\gamma}_{\rm ad} = 0$ and only radiative losses are present. When this happens, the observed slope of the steady-state non-thermal particle distribution will be $p = s+1$ for every electron Lorenz factor $\gamma$. If the jet is purely adiabatic rather than isothermal, then no re-acceleration is present and $\beta_{\rm eff}$ is the true expansion speed of the jet; this may or may not dominate over radiative cooling, depending on the particle energy. In practice, the true value of $\beta_{\rm eff}$ is somewhere between these two extremes, and allows one to fine-tune the break in the particle distribution as we did in Paper 1 (in that work, $f_{\rm b} = \beta(z)/\beta_{\rm eff}$, where $\beta(z)$ is the jet speed along the z-axis).

The calculation of the maximum electron energy is unchanged from previous versions. We define the particle acceleration time-scale as:
\begin{equation}
    t_{\rm acc} = \frac{4\gamma m_{\rm e}c}{3f_{\rm sc}eB(z)},
\end{equation}
where $e$ is the charge of the electron, $B(z)$ the magnetic field strength along the jet, and $f_{\rm sc}$ is a free parameter (described in \citealt{Jokipii87}) that quantifies the efficiency of particle acceleration, originally in terms of a relative shock velocity and ratio to scattering mean free path, but now this is grouped into a single parameter. The maximum energy of the injected particles $\gamma_{\rm max, pl}$ is found by solving:
\begin{equation}
t_{\rm acc}^{-1} = t_{\rm ad}^{-1} + t_{\rm rad}^{-1},
\end{equation}
where $t_{\rm ad}$ and $t_{\rm rad}$ are the timescales derived from equations \ref{eq:rad_dot} and \ref{eq:ad_dot} respectively. This results in a maximum injected Lorenz factor:

\begin{align}
\gamma_{\rm max, pl}(z) = & \frac{-3m_{\rm e}c^{2}\beta_{\rm eff}}{8\sigma_{\rm t}U_{\rm rad}(z) r(z)} 
\nonumber
\\ & +\frac{1}{2}\sqrt{\left(\frac{-3m_{\rm e}c^{2}\beta_{\rm eff}}{4\sigma_{\rm t}U_{\rm rad} r(z)}\right)^{2}+\frac{3f_{\rm sc}eB(z)}{4\sigma_{\rm t}U_{\rm rad}}} 
\label{eq:max_gamma}
\end{align}

\subsection{The role of pairs}

The second update to \texttt{bljet} presented in this paper consists of a more thorough calculation of the pair content of the jets; previously, \texttt{bljet} only considered jets which contain one proton per electron. Abandoning the assumption of one pair per electron requires a change in the equipartition conditions of the plasma injected at the jet base, compared to Paper 1.  Similarly to Paper 1, we define the power injected in the jet as:
\begin{equation}
N_{\rm j} = 2\gamma_0 \beta_0 c \pi R_0^{2} \left(U_{\rm b,0} + U_{\rm p,0} + U_{\rm e,0} \right),
\label{eq:Nj}
\end{equation}
where the factor 2 accounts for the launching of two jets, $\gamma_0$ is the initial jet Lorentz factor, $\beta_0$ the initial jet speed in units of $c$, $R_0$ the radius of the jet base, $U_{\rm b,0} = B^{2}/8\pi$ the energy density in magnetic fields in the jet base, $U_{\rm p,0} = n_{\rm p} m_{\rm p} c^{2}$ the energy density of the (cold) protons injected in the jet, $U_{\rm e,0} = n_{\rm e} \langle \gamma \rangle m_{\rm e} c^{2}$ the energy density of the injected (hot) electrons. We define the pair content of the jet as $\eta = n_{\rm e}/n_{\rm p}$. As in Paper 1, we wish to describe how these three components are related to each other in order to fully describe the energy budget of the jet. For convenience, in our code we compute $\eta$ as a function of the initial magnetisation $\sigma_{\rm 0}$ and initial plasma-$\beta$ parameter (defined below), although we stress that this does not imply a physical causality between these numbers. We define the initial magnetisation $\sigma_{\rm 0}$ as:
\begin{equation}
\sigma_{\rm 0} = \frac{U_{\rm b,0} + P_{\rm b,0}}{U_{\rm p,0} + U_{\rm e,0} + P_{\rm e,0}} = \frac{2 U_{\rm b,0}}{U_{\rm p,0} + \Gamma_{\rm ad} U_{\rm e,0}},
\label{eq:sig0}
\end{equation}

\begin{figure}
\includegraphics[width=0.5\textwidth]{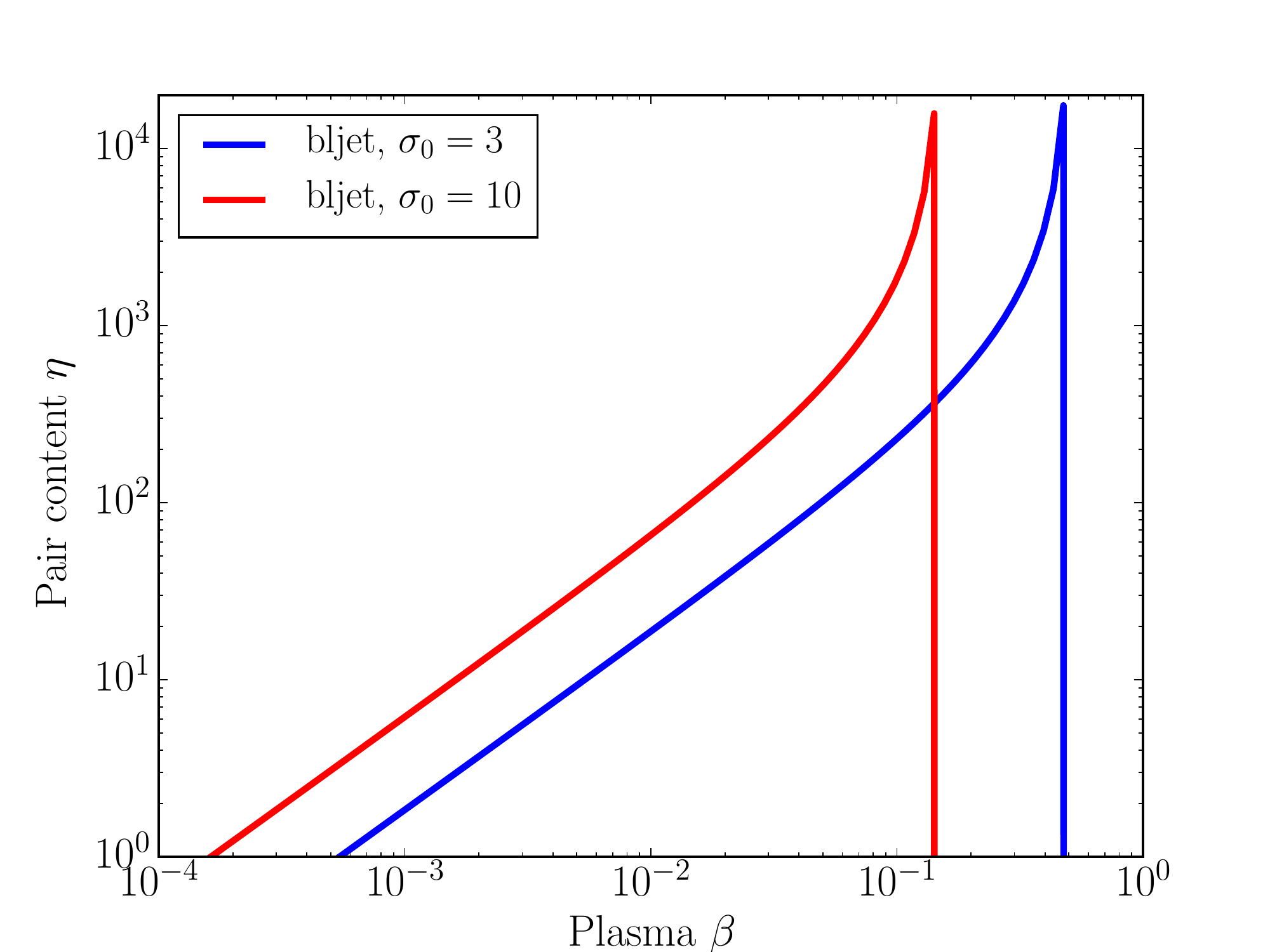}
\caption{Jet pair content as a function of initial plasma-$\beta$ $\beta_{\rm e,0}$, for both mildly and highly magnetised jets ($\sigma_{\rm 0}= 3$ and $10$, respectively), for $T_{\rm e}=135$ keV, corresponding to $\langle \gamma \rangle \approx 1.5$. } 
\label{fig:eta}
\end{figure}

where $\Gamma_{\rm ad}$ is the adiabatic index of the hot electrons; for simplicity we always take $\Gamma_{\rm ad} = 4/3$. As long as the electrons do not dominate the energy budget, this approximation (and in general, the electron population) has negligible impact on the jet dynamics. The plasma-$\beta$ parameter at the jet base is:
\begin{equation}
\beta_{\rm e,0} = \frac{U_{\rm e,0}}{U_{\rm b,0}}.
\label{eq:plasma_beta}
\end{equation}
The pair content $\eta$ is determined by the values of $\sigma_{\rm 0}$ and $\beta_{\rm e,0}$, as well as the electron average Lorentz factor $\langle \gamma \rangle$ and does not depend on any other quantities like black hole mass or injected power. This can be shown by solving equation \ref{eq:plasma_beta} for $U_{\rm b,0}$ and solving equation \ref{eq:sig0} for $\eta$:
\begin{equation}
\sigma_{\rm 0} = \frac{2 U_{\rm e,0}/\beta_{\rm e,0}}{U_{\rm p,0} + \Gamma_{\rm ad} U_{\rm e,0}} = \frac{2\eta \langle \gamma \rangle n_{\rm p} m_{\rm e}c^{2}}{\beta_{\rm e,0}\left(n_{\rm p} m_{\rm p} c^{2}+\Gamma_{\rm ad} \eta \langle \gamma \rangle n_{\rm p} m_{\rm e} c^{2}\right)},  
\end{equation}
which gives
\begin{equation}
\eta(\beta_{\rm e,0},\sigma_{\rm 0}) = \frac{\beta_{\rm e,0}\sigma_{\rm 0}}{\langle \gamma \rangle(2-\Gamma_{\rm ad} \beta_{\rm e,0}\sigma_{\rm 0})}\frac{m_{\rm p}}{m_{\rm e}}.  
\label{eq:pairs}
\end{equation}
This expression has two critical values, beyond which the the combination of $\sigma_{\rm 0}$ and $\beta_{\rm e,0}$ results in a charged, un-physical jet. The first critical value occurs when
\begin{equation}
\beta_{\rm e,0} \geq 2/(\Gamma_{\rm ad}\sigma_{\rm 0});
\end{equation}
in this case equation \ref{eq:pairs} forces the pair content of the jet to be a negative value. The second critical value is
\begin{equation}
\beta_{\rm e,0} = \frac{2 \langle \gamma \rangle}{\sigma_{\rm 0}\left(m_{\rm p}/m_{\rm e} + \langle \gamma \rangle \Gamma_{\rm ad} \sigma_{\rm 0}\right)};
\end{equation}
in this case, the jet contains exactly one electron per proton, and no pair content is necessary. Within these two critical values, the pair content of the jet for a given initial magnetisation is explicitly determined by $\beta_{\rm e,0}$. This is shown in Fig.\ref{fig:eta}; the magnetisation essentially sets the normalisation of the relation between $\eta$ and $\beta_{\rm e,0}$. We note that in the near-relativistic regime which this paper focuses on, $\langle \gamma \rangle \approx 1-3$ and therefore the impact of the electron temperature on the pair content is far smaller than that of $\beta_{\rm e,0}$ and $\sigma_{\rm 0}$. 

\begin{figure}
\includegraphics[width=0.48\textwidth]{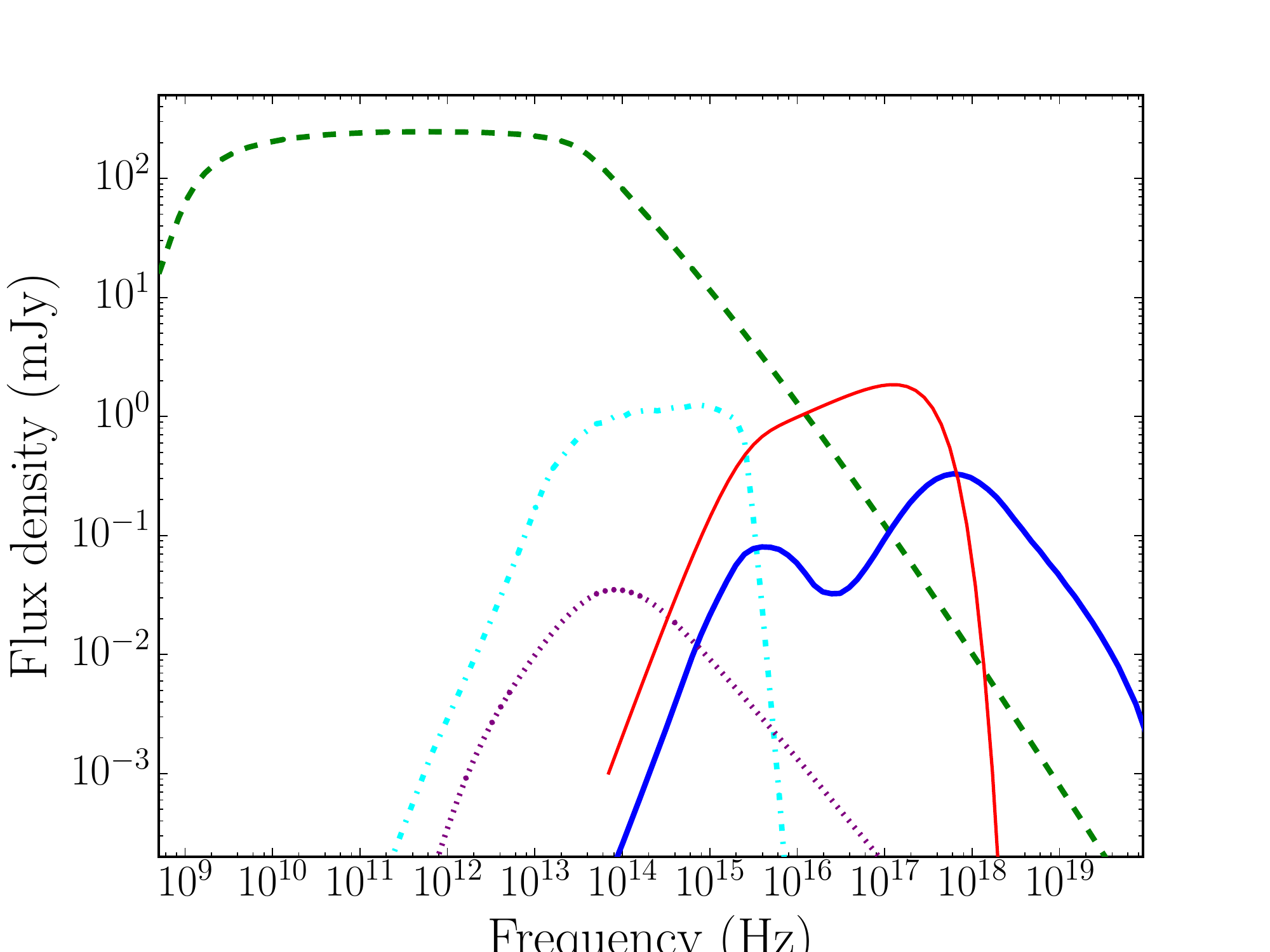}\quad
\includegraphics[width=0.48\textwidth]{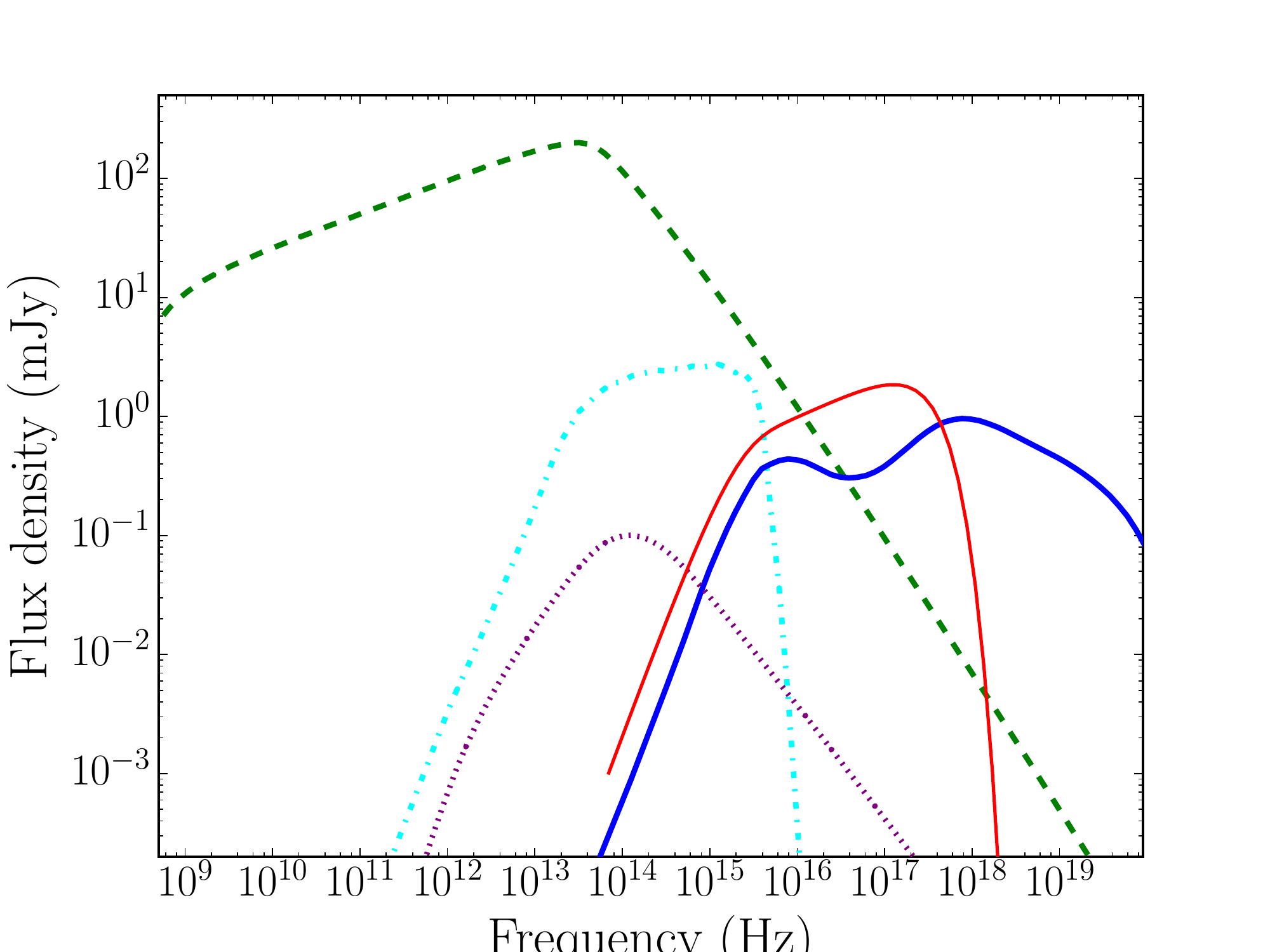}
\caption{Effect of the $f_{\rm pl}$ parameter on the radio emission. The two SEDs show the difference between jets with different powers ($N_{\rm j} = 4 \cdot 10^{-2}\,L_{\rm Edd}$ and $N_{\rm j} = 8 \cdot 10^{-2}\,L_{\rm Edd}$, respectively) and different values of  $f_{\rm pl}$ ( $f_{\rm pl} = 0$ and  $f_{\rm pl} = 10$, respectively). Both SEDs assume $M_{\rm bh} = 10\,\rm{M_{\odot}}$, $D=7\,\rm{kpc}$, $T_{\rm e} = 100\,\rm{keV}$, $r_{\rm 0}=10\,\rm{R_g}$, $z_{\rm diss}=10^{4}\,\rm{R_g}$, $L_{\rm disk} = 4\cdot 10^{-2} L_{\rm Edd}$. The dashed green and dotted purple lines represent non-thermal synchrotron and inverse-Compton emission, respectively; the dotted cyan and continuous thick blue lines represent thermal cyclo-synchrotron and inverse-Compton emission from the jet base, respectively; the continuous thin red line the emission from the disk.} 
\label{fig:pldist}
\end{figure}

\subsection{Optically thick spectral shapes}
\label{sec:radio}

The final update to \texttt{bljet} presented in this paper is a simple parametrisation designed to allow more flexibility in fitting optically-thick spectra from compact jets. 

The radio spectrum of \MAXI J1836-194 was highly inverted rather than slightly-inverted/flat throughout its outburst \citep{Russell15}, with spectral indexes ranging from $\alpha = 0.19$ to $\alpha = 0.70$ (where $F(\nu) \propto \nu^{\alpha}$). This behaviour cannot be captured by the simple assumption made in the last version of \texttt{bljet} of a conical, isothermal jet, which produces a strictly flat radio spectrum ($F(\nu) \propto \nu^{0}$, \citealt{Blandford79}), or in the quasi-isothermal treatment in \texttt{agnjet} \citep[see, e.g.][]{Crumley17}. The optically-thick slope is set by the details of the balance between particle acceleration and cooling. \texttt{bljet} does not capture these details self-consistently, as this would require a full Fokker-Planck treatment linking particle acceleration and cooling to the jet geometry and dynamics.; therefore, we introduced a phenomenological free parameter in the particle distribution to gradually reduce the emissivity along the jet z-axis. We produce inverted radio spectra by changing the fraction of electrons channelled in the non-thermal tail along the length of the jet:
\begin{equation}
n_{\rm nth}(z) = n_{\rm nth,0}\left(\frac{\log_{\rm 10}(z_{\rm diss})}{\log_{\rm 10}(z)}\right)^{f_{\rm pl}}
\label{eq:pldist}
\end{equation}

where $n_{\rm nth,0} = 0.1$ is the fraction of particles channelled into the non-thermal tail at $z_{\rm diss}$, $z$ is the distance from the black hole, and $f_{\rm pl}$ is a free parameter which we fit to the data in order to match the optically thick spectral shape. Higher values of $f_{\rm pl}$ suppress the number density of non-thermal particles, mimicking the effect of additional adiabatic cooling and/or reduced particle acceleration, that can hopefully guide the implementation of more physical treatments in future work.

A limitation of this phenomenological approach comes from estimating the jet power by modelling exclusively the optically-thick radio flux densities. This is highlighted in Fig.\ref{fig:pldist}. Suppressing the emission from the outer jet in order to invert the radio spectrum can lead to a counter-intuitive regime: due to the shape of the optically-thick spectrum, a less powerful, flat-spectrum jet can produce more radio emission than a more powerful, inverted-spectrum one. A similar behaviour would also emerge from a more self-consistent treatment, in which the radio spectrum is inverted with an additional loss term in the Fokker-Planck equation. Either treatment essentially quantifies how much of the jet initial power is converted in radio emission. The goal of this paper is to first quantify this behaviour, before implementing a more self-consistent treatment in future works.

\subsection{Model parameters}
\label{sec:pairs}

The focus of this work is to probe the connection between the accretion disk, jet base, and outer compact jet as the source evolves through its outburst; as such, we froze parameters that can be estimated by observations (such as black hole mass), do not greatly impact the SED (such as the location of the cooling break or the outer disk radius), or introduce large amounts of degeneracy (such as the pair content). Furthermore, we fixed the value of \nH to that found in section \ref{sec:Xrayfits}.

We assume a black hole mass of 10 $M_{\odot}$, a distance of 7 kpc, and a viewing angle of $10^{\circ}$ (\citealt{Russell14a}, \citealt{Russell14b}). 

\begin{table}
\centering
\begin{tabular}{@{}cp{5.5cm}}
\hline \hline
Fixed Parameter & Description \bigstrut \\ \hline
$M_{\rm bh} = 10 M_{\odot}$ & Mass of the black hole \bigstrut \\ \hline
$D = 7$ kpc & Distance to the source \bigstrut \\ \hline
$\theta = 10^{\circ}$ & Source inclination \bigstrut \\ \hline
$\gamma_{\rm f} = 3$ & Terminal jet Lorenz factor after bulk acceleration \bigstrut \\ \hline
$\sigma_{\rm f} = 0.1$ & Final jet magnetisation after bulk acceleration \bigstrut \\ \hline
$f_{\rm sc} = 0.1$ & Particle acceleration efficiency \bigstrut \\ \hline
$\beta_{\rm eff} = 0.1$ & Adiabatic cooling efficiency \bigstrut \\ \hline
$\beta_{\rm e,0} = 0.02085$ & Initial plasma-$\beta$ at the jet base; sets the pair content. Corresponds to $\approx 20$ pairs per proton with $\gamma_{\rm f} = 3$, $T_{\rm e}\approx 100\,\rm{keV}$ and $\sigma_{\rm 0} \approx 2$  \bigstrut \\ \hline
$f_{\rm heat} = 1$ & Shock heating parameter \bigstrut \\ \hline
$R_{\rm out} = 10^{5} R_{\rm g}$ & Disk outer radius \bigstrut \\ \hline\hline
Fitted Parameter & Description \bigstrut \\ \hline
$\boldsymbol{N_{\rm j}}$ ($L_{\rm Edd}$) & Power injected in the jet base \bigstrut \\ \hline
$\boldsymbol{r_{\rm 0}}$ (${\rm R_g}$) & Jet base radius \bigstrut \\ \hline
$\boldsymbol{z_{\rm diss}}$ (${\rm R_g}$) & Location where particle acceleration in the jet starts \bigstrut \\ \hline
$\boldsymbol{T_{\rm e}}$ (keV) & Electron temperature at the base of the jet \bigstrut \\ \hline 
$\boldsymbol{f_{\rm pl}}$ & Additional loss parameter to set the radio spectral index \bigstrut \\ \hline
$\boldsymbol{s}$ & Slope of the injected non-thermal particle distribution \bigstrut \\ \hline
$\boldsymbol{L_{\rm disk}}$ ($L_{\rm Edd}$) & Disk luminosity \bigstrut \\ \hline
$\boldsymbol{r_{\rm in}}$ (${\rm R_g}$) & Disk truncation radius \bigstrut \\ \hline
$\boldsymbol{T_{\rm bb}}$ (K) & Optical excess black body temperature \bigstrut \\ \hline
$\boldsymbol{L_{\rm bb}}$ ($\rm{erg s^{-1}}$) & Optical excess black body luminosity \bigstrut \\ \hline\hline
\end{tabular}
\caption{Summary of the parameters of \texttt{bljet}. The parameters in the top block are frozen, as they can be estimated a-priori or do not impact the SED. The parameters in the bottom block, shown in bold font, are kept free and fitted to the data.}
\label{tab:pars}
\end{table}
\begin{figure*}
\includegraphics[width=0.48\textwidth]{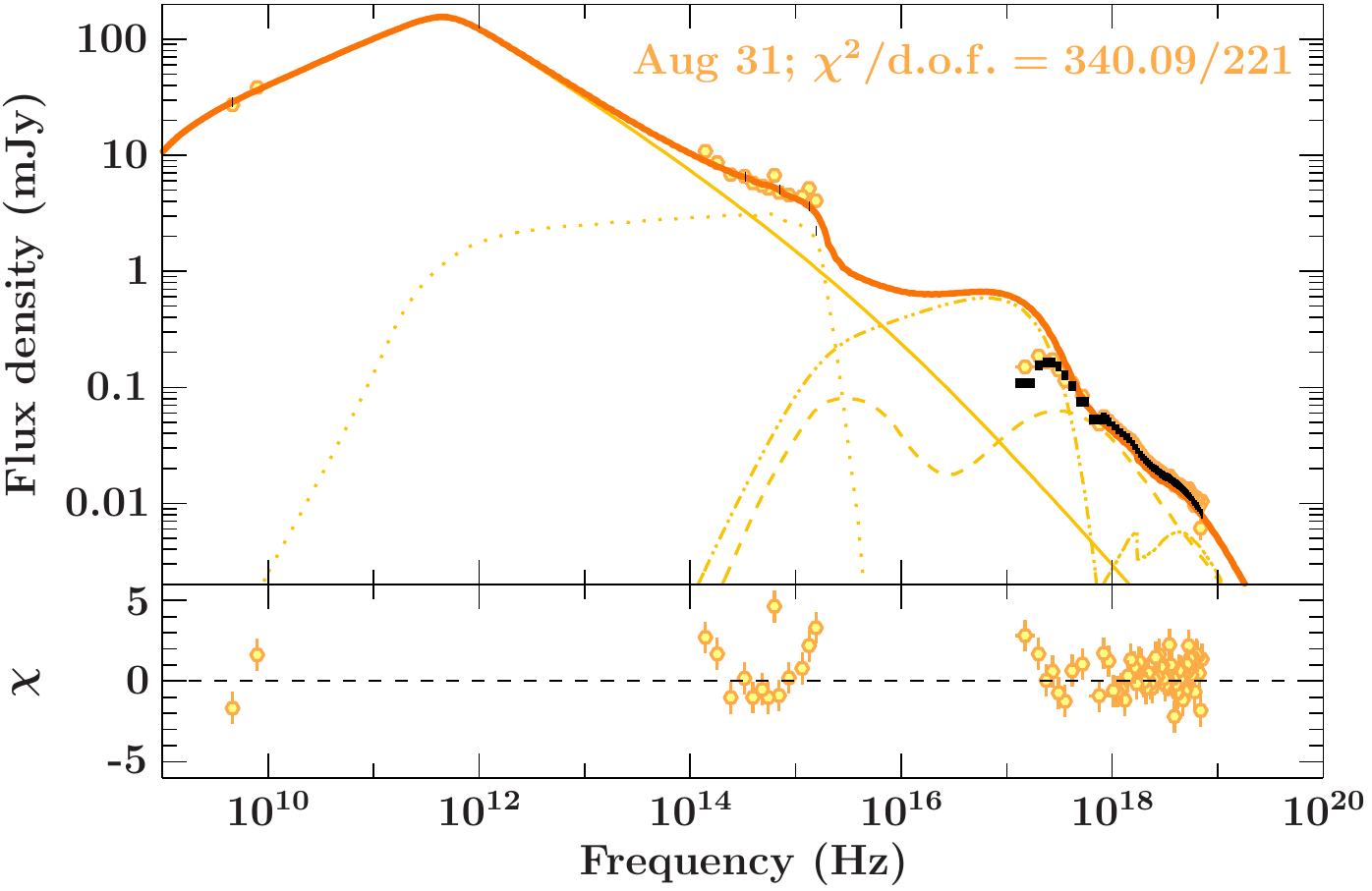}\quad
\includegraphics[width=0.48\textwidth]{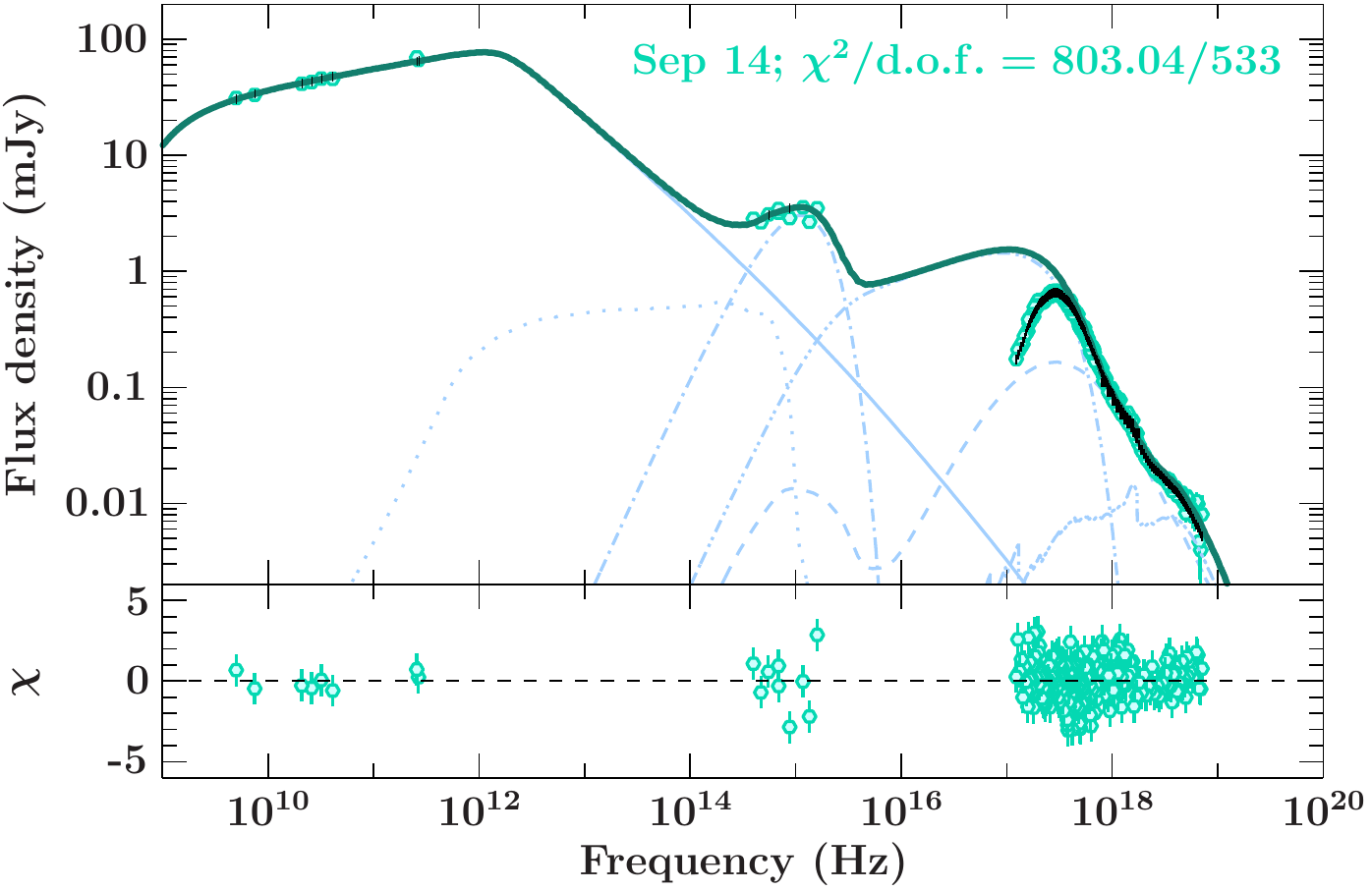}
\includegraphics[width=0.48\textwidth]{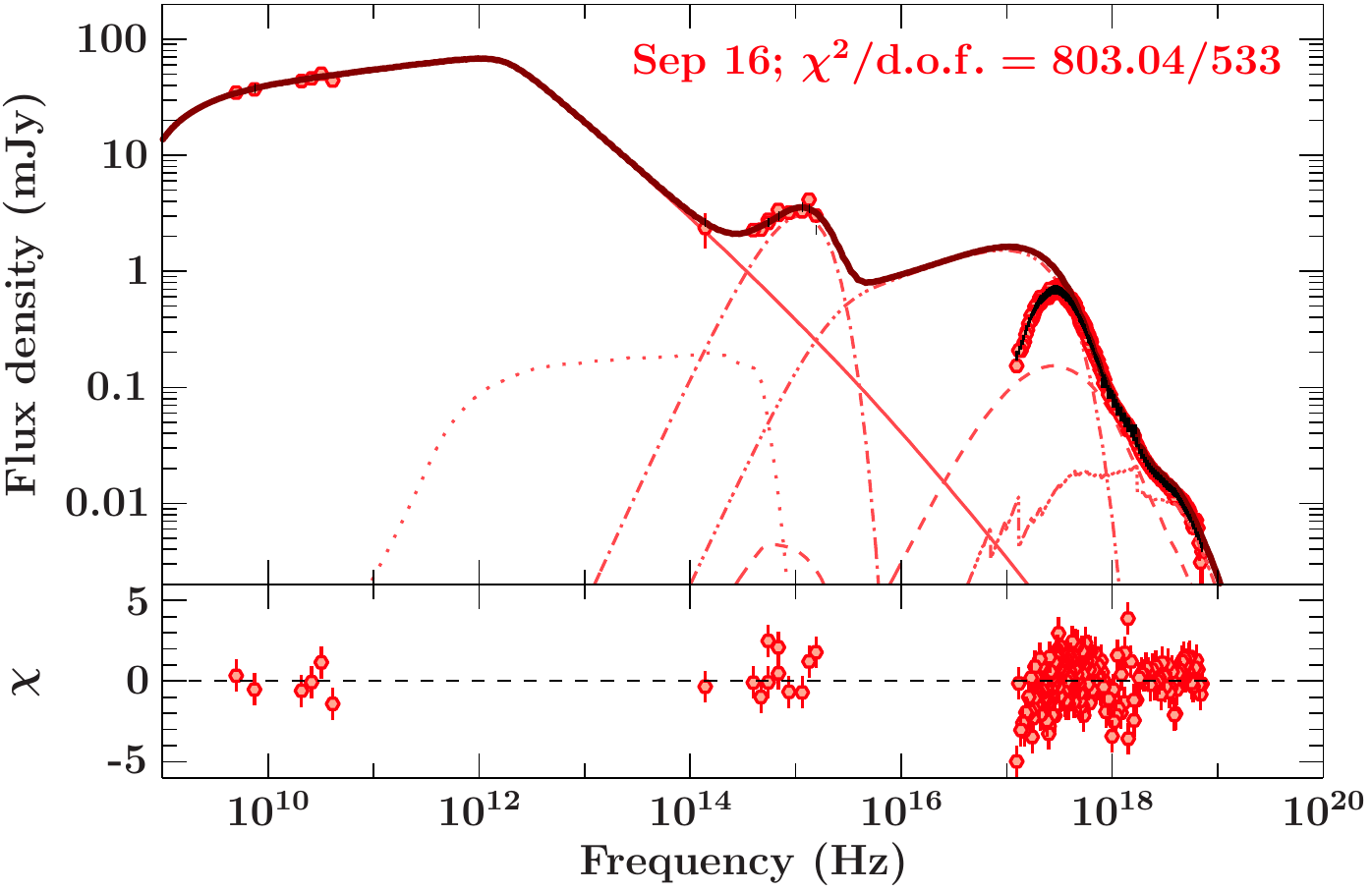}\quad
\includegraphics[width=0.48\textwidth]{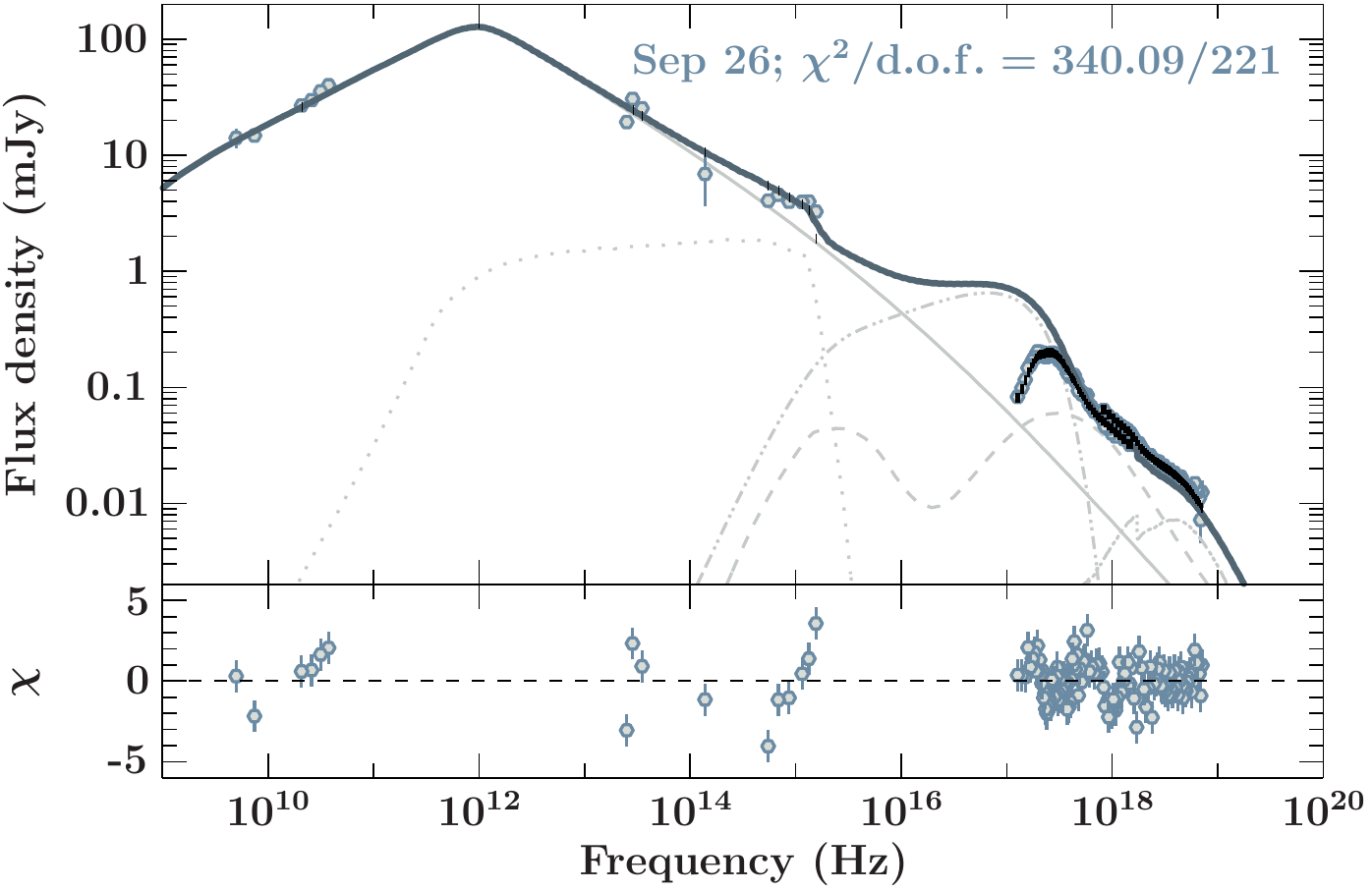}
\includegraphics[width=0.48\textwidth]{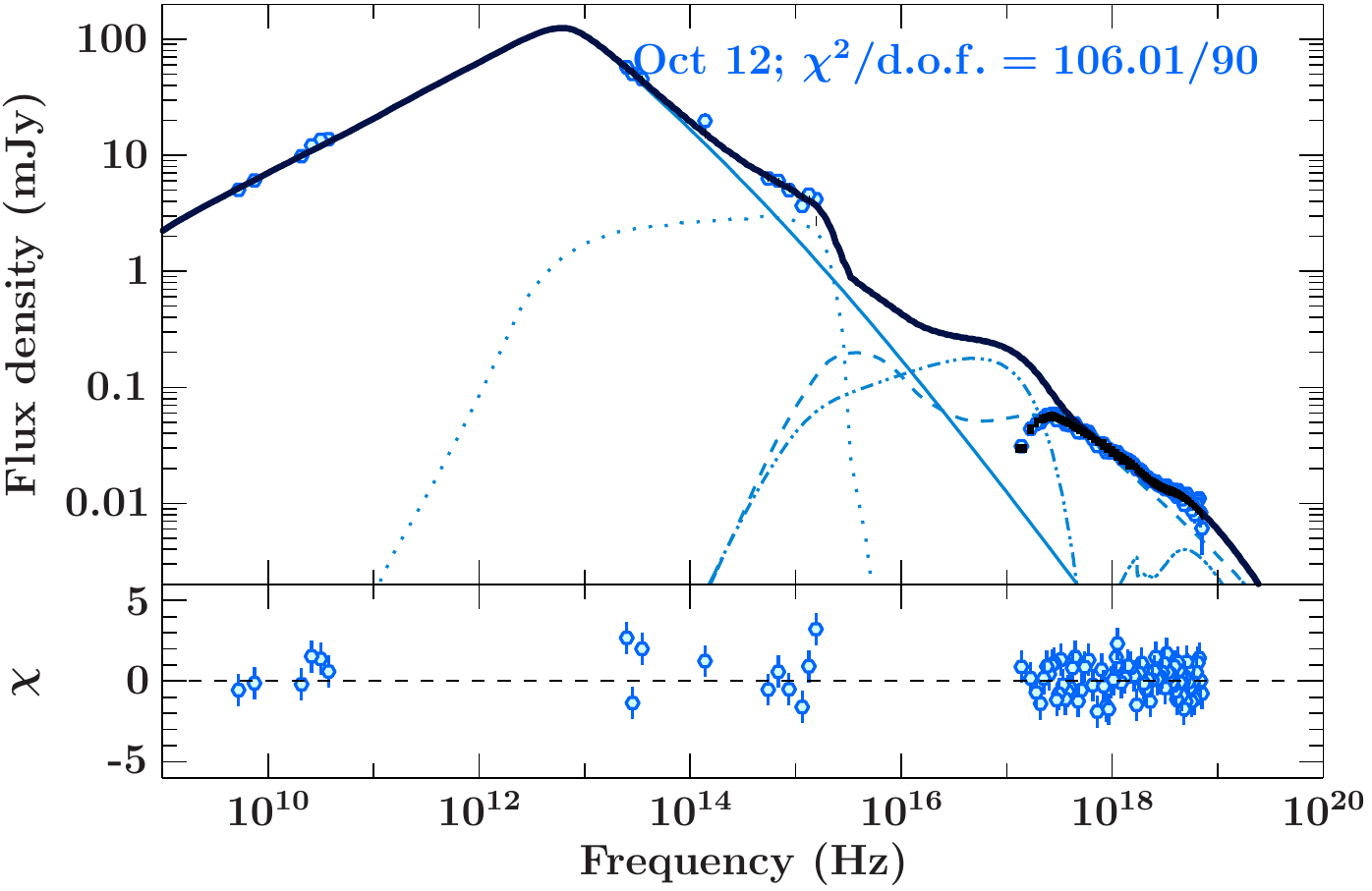}\quad
\includegraphics[width=0.48\textwidth]{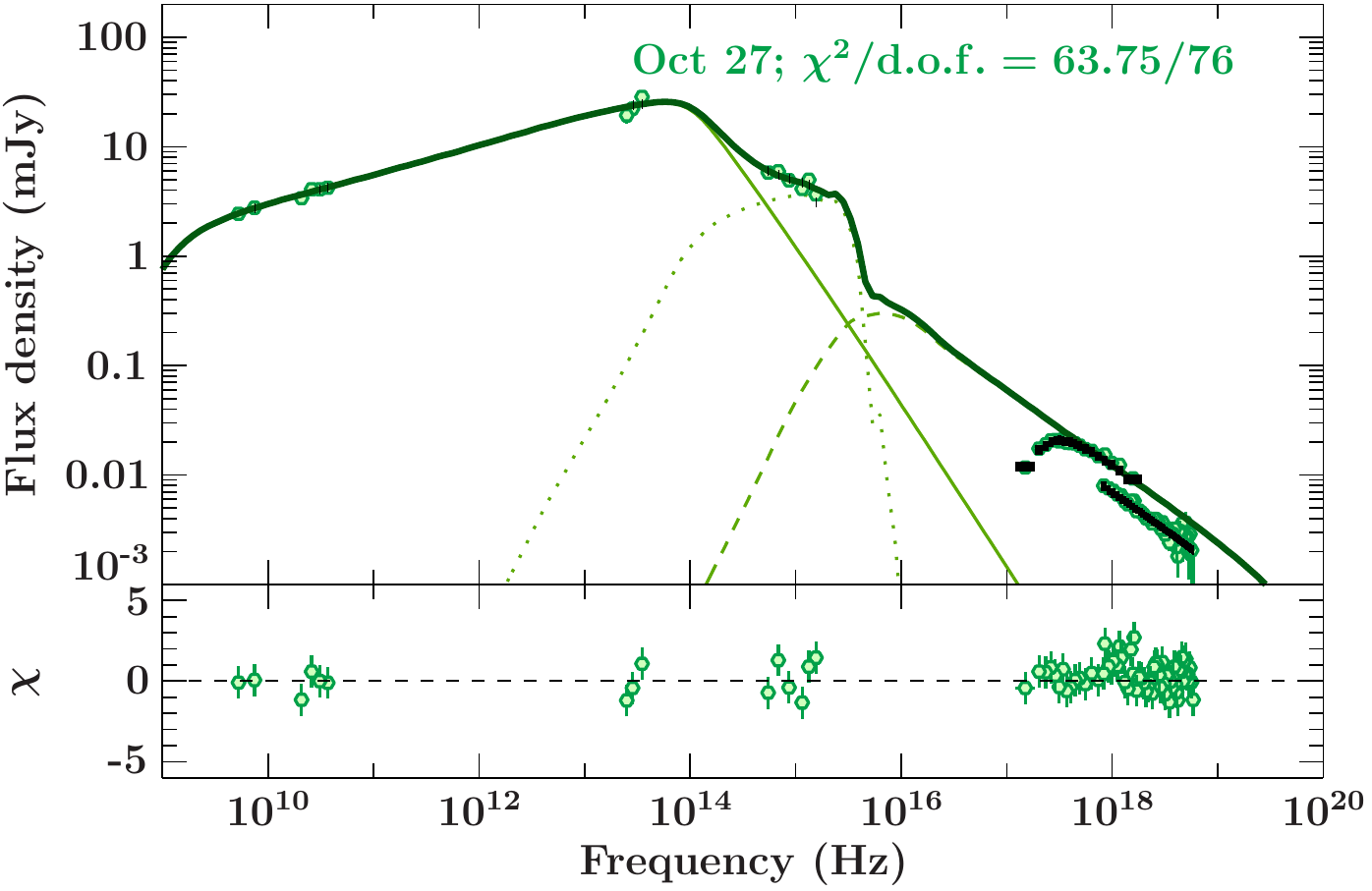}
\label{fig:seds}
\caption{Best fit of each SED; August 30 and September 26, and September 13 and 16 have been fitted jointly, while the October epochs are fit individually. The thick continuous dark line represents the total model, the thin continuous line represents the non-thermal synchrotron emission, the thin dashed line the inverse-Compton emission from the jet base, the thin dotted line represents the thermal cyclo-synchrotron emission from the jet base, the dot-dashed and double dot-dashed lines represent the optical black body excess and accretion disk, the triple dot-dashed line represents reflection.} 
\end{figure*}

X-ray binary jets are believed to be only mildly relativistic, thus we take $\gamma_{\rm f} = 3$ \citep{Fender04a}. Constraining the value of $\sigma_{\rm f}$ is harder due to modelling degeneracies, particularly in the case of X-ray binary jets in which the non-thermal inverse Compton emission of the jets is not detected. Theoretical models however predict that the magnetisation will continue  to decrease as the jet accelerates, until the jet either reaches equipartition or is slightly matter-dominated (e.g. \citealt{Tchekhovskoy09}, \citealt{Komissarov09}, \citealt{Ceccobello18}, \citealt{Chatterjee19}). With some exceptions, blazar jets also appear to either be close to equipartition or be somewhat matter dominated (e.g. \citealt{Ghisellini14}, Paper 1). Therefore, we fix $\sigma_{\rm f}=0.1$ in all of our fits.
 
Similarly to \cite{Lucchini19b}, but unlike Paper 1, we find that no additional heating of the electrons is necessary to match the data. Therefore, we fix $f_{\rm heat} = 1$. In practice, this means that the minimum Lorenz factor of the non-thermal electron distribution is $\gamma_{\rm min, pl} \approx 1$. The maximum and break energies of the electron distribution are not well constrained. As such, we take $f_{\rm sc} = 0.1$ and $\beta_{\rm eff} = 0.1$. This results in $\gamma_{\rm max, pl} \approx 10^{6}$ along the jet, varying by a factor of $\approx$ a few with the distance from the black hole. Physically, this means we do not make any a-priori assumption on the radiative mechanism responsible for the X-ray emission, and allow the non-thermal synchrotron emission to extend above the \RXTE/PCA band.

\begin{table*}
\caption{Summary of the best-fitting parameters for each fit (individual for the October data, jointly for the other epochs), using \texttt{bljet} for the continuum.}
\begin{tabular}{lcccccc}
Parameter &  Aug 30 & Sep 13 & Sep 16 & Sep 26 & Oct12 & Oct 27\\ \hline
$N_{\rm j}$ ($10^{-2}$ L$_{\rm Edd}$)& $4.3^{+0.3}_{-0.4}$ & $4.2^{+0.3}_{-0.2}$ & $3.6^{+0.3}_{-0.3}$ & $3.7^{+0.3}_{-0
.3}$ & $7.7^{+1.1}_{-1.1}$ & $6.2^{+2.4}_{-2.2}$ \\
$r_{\rm 0}$ (R$_{\rm g}$)$^{a}$ & $18.6^{+1.4}_{-1.3}$ & $25.3^{+1.0}_{-1.1}$ & $25.3^{+1.0}_{-1.1}$ & $18.6^{+1.4}_{-1.3}$ & $17.5^{+2.5}_{-2.6}$ & $12.0^{+4.5}_{-3.7}$  \\
$z_{\rm diss}$ (R$_{\rm g}$) & $7.4^{+1.8}_{-2.6}\cdot10^{5}$ & $1.8^{+0.9}_{-0.5}\cdot 10^{5}$ & $1.5^{+0.9}_{-0.6}\cdot 10^{5}$ & $4.0^{+0.5}_{-0.5}\cdot 10^{5}$ & $6.6^{+1.2}_{-1.1}\cdot 10^{4}$ & $2.5^{+0.7}_{-0.5}\cdot 10^{3}$ \\ 
$T_{\rm e}$ (keV) & $126^{+4}_{-4}$ & $70^{+4}_{-5}$ & $52^{+10}_{-7}$ & $114^{+7}_{-7}$ & $101^{+5}_{-3}$ & $126^{+18}_{-8}$\\ 
$f_{\rm pl}$ & $15.3^{+1.5}_{-2.9}$ & $5.0^{+0.7}_{-0.7}$ & $3.4^{+0.9}_{-1.0}$ & $19.5^{+1.0}_{-0.7}$ & $18.1^{+1.3}_{-1.3}$ & $7.5^{+0.5}_{-0.5}$ \\
$s$  & $2.15^{+0.08}_{-0.06}$  & $2.55^{+0.10}_{-0.05}$ & $2.55^{+0.06}_{-0.07}$ & $2.10^{+0.04}_{-0.04}$ & $2.45^{+0.06}_{-0.07}$ & $3.0^{+0.2}_{-0.2}$ \\
$L_{\rm disk}$ ($10^{-2}$ L$_{\rm Edd}$) & $0.64^{+0.04}_{-0.03}$ & $2.27^{+0.02}_{-0.02}$ & $2.44^{+0.04}_{-0.03}$ & $0.74^{+0.02}_{-0.01}$ & $0.15^{+0.01}_{-0.01}$ & // \\
$r_{\rm in}$ (R$_{\rm g}$)$^{a}$ & $15.6^{+0.5}_{-0.4}$ & $14.0^{+0.2}_{-0.1}$ & $14.0^{+0.2}_{-0.1}$ & $15.6^{+0.5}_{-0.4}$ & $ 12.8^{+1.9}_{-1.6}$ & // \\
$L_{\rm bb}$ ($10^{36}\,\rm{erg s^{-1}}$)$^{a}$ & // & $1.0^{+0.2}_{-0.1}$ & $1.0^{+0.2}_{-0.1}$ & // & // & // \\
$T_{\rm bb}$ ($10^{4}\,\rm{K}$)$^{a}$ & // & $1.9^{+0.2}_{-0.1}$ & $1.9^{+0.2}_{-0.1}$ & // & // & // \\
relfrac & $1.05^{+0.11}_{-0.10}$ & $1.6^{+0.2}_{-0.2}$ & $3.8^{+0.7}_{-0.8}$ & $1.4^{+0.2}_{-0.1}$ & $0.6^{+0.1}_{-0.1}$ & //\\
\texttt{Gaussian} norm ($10^{-4}$) & $2.8^{+2.2}_{-1.6}$ & $9.7^{+3.5}_{-2.9}$ & $22^{+7}_{-9}$ & $25^{+10}_{-10}$ & $12.0^{+6.0}_{-5.0}$ & // \\
\texttt{Gaussian} $\sigma$ (keV) & $1.0^{+1.6}_{-0.7}$ & $0.4^{+0.3}_{-0.2}$ & $2.1^{+0.4}_{-0.4}$ & $4.6^{+1.3}_{-0.9}$ & $2.0^{+0.8}_{-0.6}$ & // \\ \hline
\end{tabular}
\begin{flushleft} 
\vspace{-2mm}
$^{a}$: Tied over Aug30/Sep26 and/or Sep13/Sep16.
\end{flushleft}
\label{tab:sedfits}
\end{table*}

The pair content has a large effect on the SED, changing its normalisation and the relative importance of Comptonisation. Unfortunately, this effect is almost entirely degenerate with the effect of the injected jet power. As such, we set a pair content of $\approx 20$ pairs per proton (changing slightly depending on temperature), giving $\beta_{\rm e,0} = 0.02085$. This results in jet powers roughly on the order of the accretion luminosity; on the other hand, taking one electron per proton requires unreasonable values of the injected jet power. We discuss this choice further in Section \ref{sec:discussion}.

Assuming a standard Shakura-Sunyaev temperature profile $T(R) \propto R^{-3/4}$, the emission from the outer disk is sub-dominant at every wavelength, hence we always take $R_{\rm out} = 10^{5}\,\rm{R_g}$. However, on September 13 and 16 (near the peak of the outburst) we found in preliminary fits that the spectral shape and luminosity, particularly in the UV, can not be matched by cyclo-synchrotron emission from the jet base. Therefore, we include an additional black body peaking at optical/UV wavelengths, likely due to disk irradiation. This is analogous to the models in \citealt{Russell14b} and \citealt{Peault19}. In every other epoch this contribution is negligible compared to the jet, and thus we choose to neglect it to reduce the number of free parameters.

In summary, our disk+jet model has up to 13 free parameters: 6 relating to the jet, 4 to the accretion flow, and 3 to the reflection component. These are: the injected jet power $N_{\rm j}$, the jet base radius $r_{\rm 0}$, the temperature of the electrons at the base $T_{\rm e}$ (which in our model can be constrained from the optical cyclo/synchrotron bump as well as the X-ray spectrum), the location of particle acceleration $z_{\rm diss}$, the slope of the accelerated power-law $s$, the (radio) spectral shape parameter $f_{\rm pl}$, the disk luminosity $L_{\rm disk}$, the disk inner radius $r_{\rm in}$, the black body luminosity $L_{\rm bb}$, the black body temperature $T_{\rm bb}$, the reflection fraction, and the normalisation and with of the Gaussian. The additional black body is only present on September 13 and 16, and the accretion disk and reflection components are absent on October 27.

\subsection{Joint fits}

We initially modelled each SED independently; while the model reproduced the data very well, we found significant degeneracies in the posterior distributions of the parameters. In order to address these issues we combined timing and spectral information, and performed joint fits of SEDs that had similar power-colours, using an approach similar to \cite{Connors19}. As shown in Fig.\ref{fig:timing}, these are September 13/14 and 16, around the peak of the outburst, and August 30/31 and September 26, close to the HS to HIMS and HIMS to HS transitions, respectively. For each of these epochs we take the geometrical parameters of the X-ray emitting regions to be the same; in the case of our model, this means tying the disk truncation radius $r_{\rm in}$ and the jet base radius $r_{\rm 0}$. The best fitting results for the joint (on August 30 and September 26, and September 13 and 16) and individual (October 12 and October 27) fits are reported in Tab.\ref{tab:sedfits}, and all the SEDs are shown in Fig.\ref{fig:seds}. In every epoch we find that the model is in excellent agreement with the data.

\section{Discussion}
\label{sec:discussion}

The main result of our modelling is that in every epoch the full broad-band SED can be fitted by the disk+jet emission, without the need to invoke contributions from an additional coronal region as in \cite{Connors19}. The reason for our better fits is our model's new ability to treat sub- and near-relativistic temperatures for the electrons at the jet base. This demonstrates that the coronal emission in these observations can in principle be closely related to the base of the jets, provided that the emitting particle distribution remains in the mildly-relativistic regime as might be expected from the inner hot accretion flow, and the optical depth is large enough to produce hard Comptonisation spectra. In all of our models we find that $\tau \approx 0.1-1$ (depending on the values of $r_{\rm 0}$ and $N_{\rm j}$), resulting in $\Gamma \approx 1.7-2.8$. Most importantly, we find that epochs with similar X-ray power spectra can be fit with similar geometries for the emitting region.

During most epochs of the outburst, the main radiative mechanism responsible for the X-ray emission is inverse-Compton scattering of disk photons in the jet base; however, this slowly evolves during the latter two epochs of the outburst, which are harder and fainter than the first four. For October 12 we find that inverse Comptonisation of cyclo-synchrotron photons produced in the jet base (SSC) accounts for $\approx 10\%$ of the coronal continuum; by October 27 the disk is not detected in the soft X-ray anymore, and as its luminosity increases SSC becomes the main channel of Comptonisation. Near the peak of the outburst in mid-September, we find very strong reflection features, which dominate over the continuum; in other epochs (again, with the exception of October 27) reflection is still present, but much weaker. We find the extent to which reflection dominates over the continuum at energies $\geq 10\,\rm{keV}$ does not depend on the continuum model chosen, and is observed both in the phenomenological fits using \texttt{nthcomp}, and in the broadband fits using \texttt{bljet}. This is in agreement with the more sophisticated reflection modelling in \cite{Reis12} and \cite{Dong20}.

\begin{figure*}
\hspace{-0.5cm}
\includegraphics[width=1.0\textwidth]{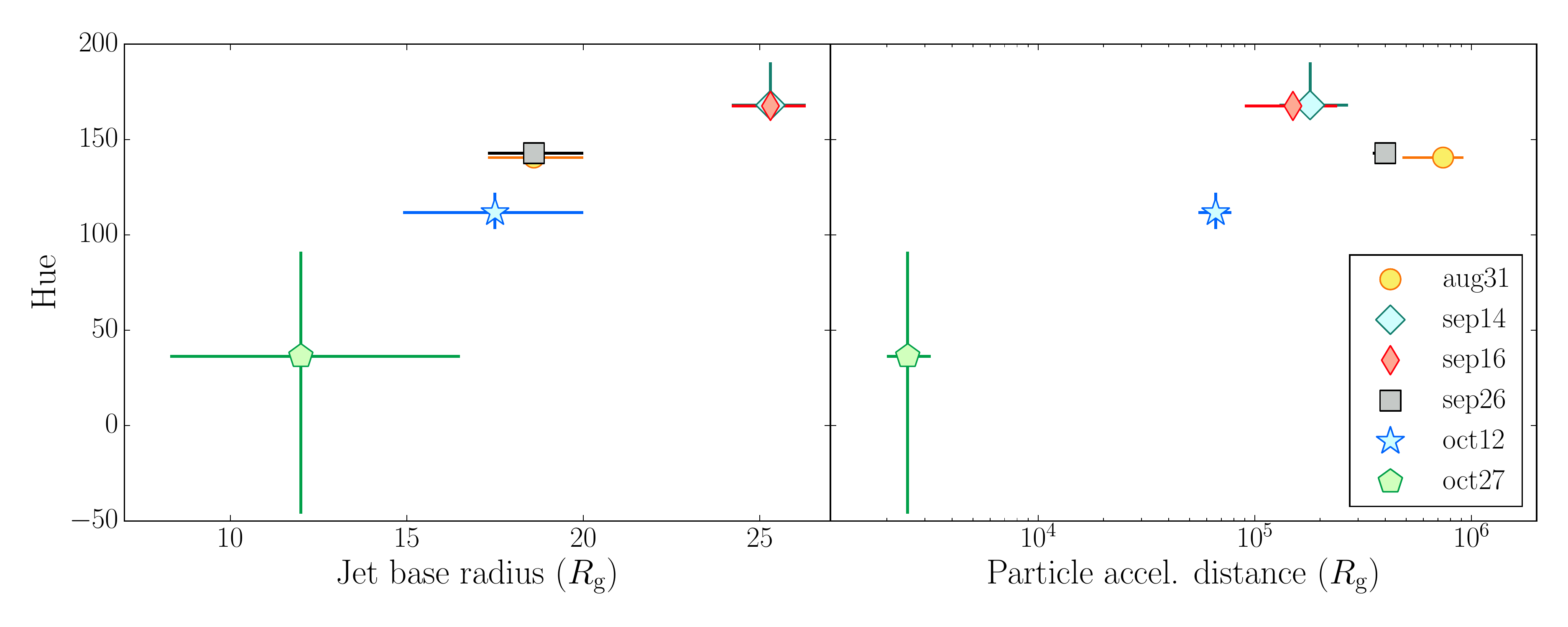}
\caption{Evolution in the jet base radius (left panel) and location of the start of particle acceleration (right panel) with power spectral hue, from our broadband fits.} 
\label{fig:parcors}
\end{figure*}

In the optical/ultraviolet bands we also find that the radiative mechanism changes during the outburst. During the HSs in October, and the HS/HIMS transitions states on August 30 and September 26, we find that the data can be represented fairly well by cyclo-synchrotron emission originating in the jet base. However, this mechanism under-predicts the UV flux of the source during the two HIMS states at the peak of the outburst. Furthermore, during these epochs the optical/UV data showed an upturn, becoming brighter at higher frequencies, which our model can not easily reproduce. This is in agreement with \cite{Russell14b}.

Finally, our fits seem to indicate that if a disk component is detected, its innermost radius is fairly constant ($R_{\rm in }\approx 12-15\,R_{\rm g}$). However, this is not a robust claim and should be taken with care. \cite{Russell14b}, using the \texttt{diskir} model, found that during the HIMS epochs the truncation radius varies with hardness between $R_{\rm in} \approx 6$ and $20\,R_{\rm g}$. Because \texttt{diskir} self-consistently accounts for disk irradiation, which our model does not, this is a more reliable estimate of the disk truncation radius. 

\subsection{An evolving jet during the outburst}

Fig.\ref{fig:parcors} shows the evolution of the jet dynamical properties, defined by the jet initial radius $r_{\rm 0}$ and the particle acceleration distance $z_{\rm diss}$, throughout the outburst, as a function of power-spectral hue (defined as the angle between the coloured and blue-grey lines in the left panel of Fig.\ref{fig:timing}). Intriguingly, both parameters appear to be correlated with the hue (although we note that in case of $z_{\rm diss}$, the appearent correlation is weak and largely driven by including the Oct 27th data-set), which is a tracer of X-ray variability properties. We stress that the hue captures the shape of the power spectrum \citep{Heil15a}, rather than a simply ratio of power at high vs low frequency: values near $0^{\circ}$ represent a flat power spectrum, while values closer to $180^{\circ}$ correspond to a more peaked power spectrum. This makes the hue a useful tool to categorise spectral states, but means it is not straightforward to deduce information about the characteristic size of the emitting region. 

The changes in jet base radius can be understood in terms of optical depth. As the outburst progresses from the HS to the HIMS, the power-law continuum softens. As the temperature in the jet base is only changing by a factor of $\approx 2$ at most (see Tab.\ref{tab:sedfits}), one natural way to soften the Comptonisation spectrum is to reduce the optical depth by increasing the size of the emitting region. Such a trend is in broad agreement with recent GRMHD simulations of thin disks. \cite{Liska19} found that compared to the standard thick disk scenario, a thin disk can still launch and collimate a powerful jet; however, compared to a thick ADAF-type disk, a thin disk exerts a lower pressure on the jet edge, leading to a wider outflow. A change in the initial jet collimation would naturally lead to the behaviour we found in our fits, where harder, fainter states with low hue (presumably close to a higher scale-height ADAF-type disk) favour smaller jet bases, while bright intermediate states with large hue require more extended emitting regions. We also note that \cite{Dong20}, who modelled the reflection spectra of the source near the peak of the outburst, favour an extended corona over a compact lamp-post geometry, in agreement with our scenario. 

Interestingly, the trend we observe of the corona/jet launching becoming wider (and more optically thin) appears to be unlike what is observed in other sources. \cite{Kara19} found that the reverberation lags in \MAXI\,J1820$+$070 become shorter as the source transitioned towards the soft state, while the iron line remained unchanged (and broad); they interpret this behaviour as the corona contracting during the state transition. \cite{Vincentelli19} found that in the HS of GX\,339$-$4 the IR power spectra evolved more slowly than the X-ray power power spectra, suggesting that the size of the jet launching region remains constant as the source evolves during the HS. However, we (also) notice that these results were found (with different methods and) during different phases of the outbursts, therefore a quantitative comparison between these estimates is not straightforward.

\begin{figure*}
\includegraphics[width=0.48\textwidth]{Plots/Pairs_lowNj.pdf}
\includegraphics[width=0.48\textwidth]{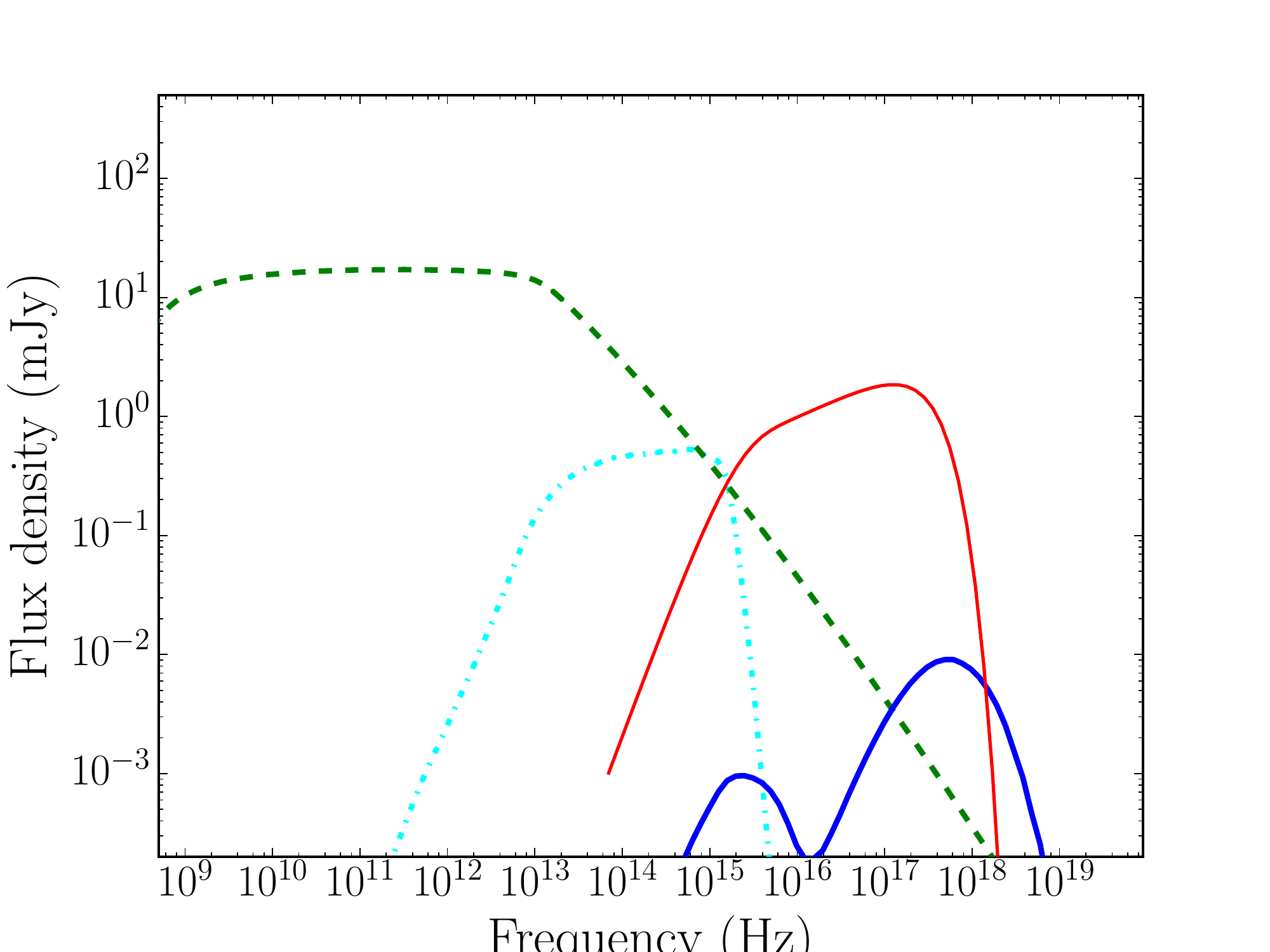}
\includegraphics[width=0.48\textwidth]{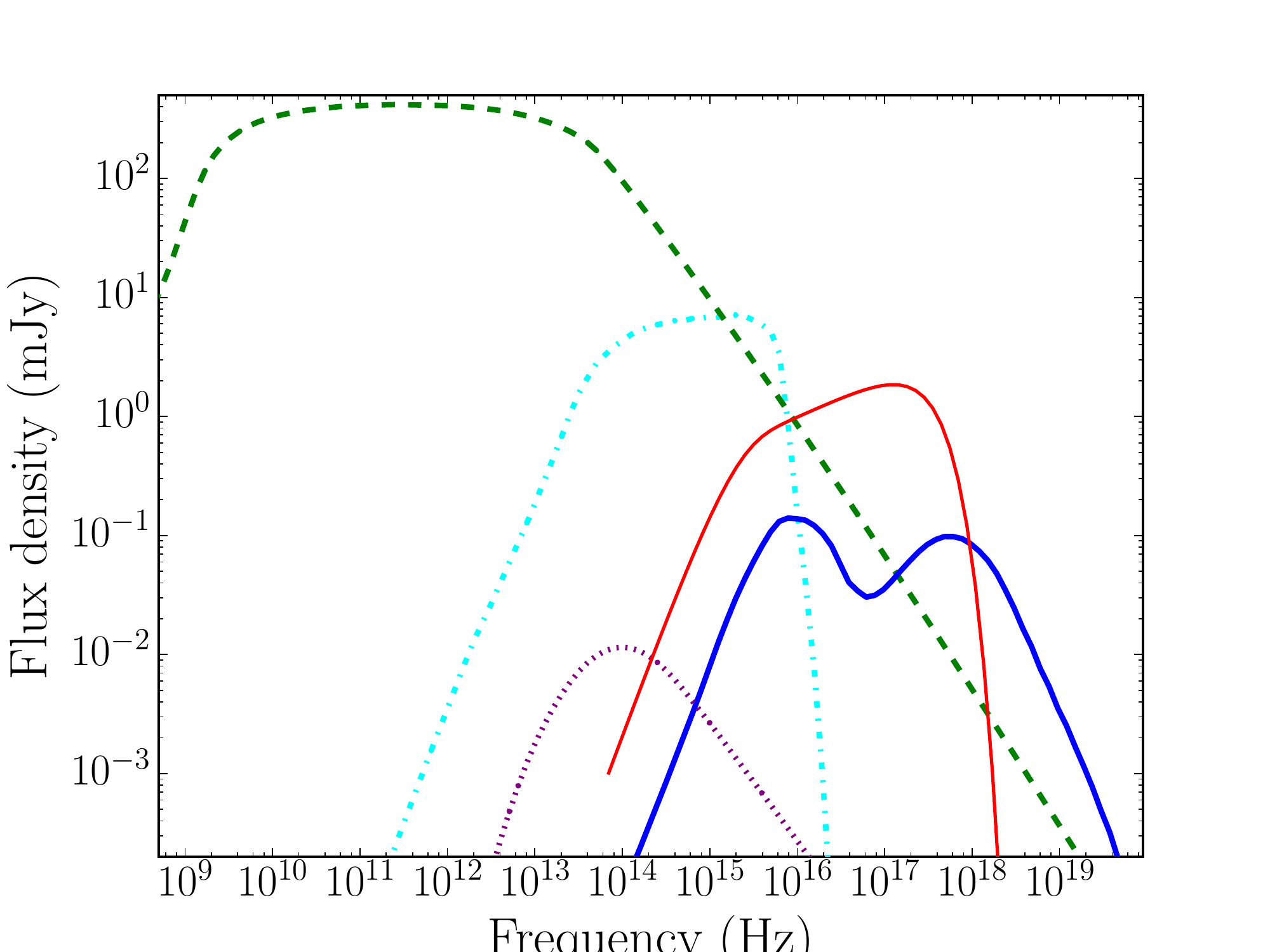}
\includegraphics[width=0.48\textwidth]{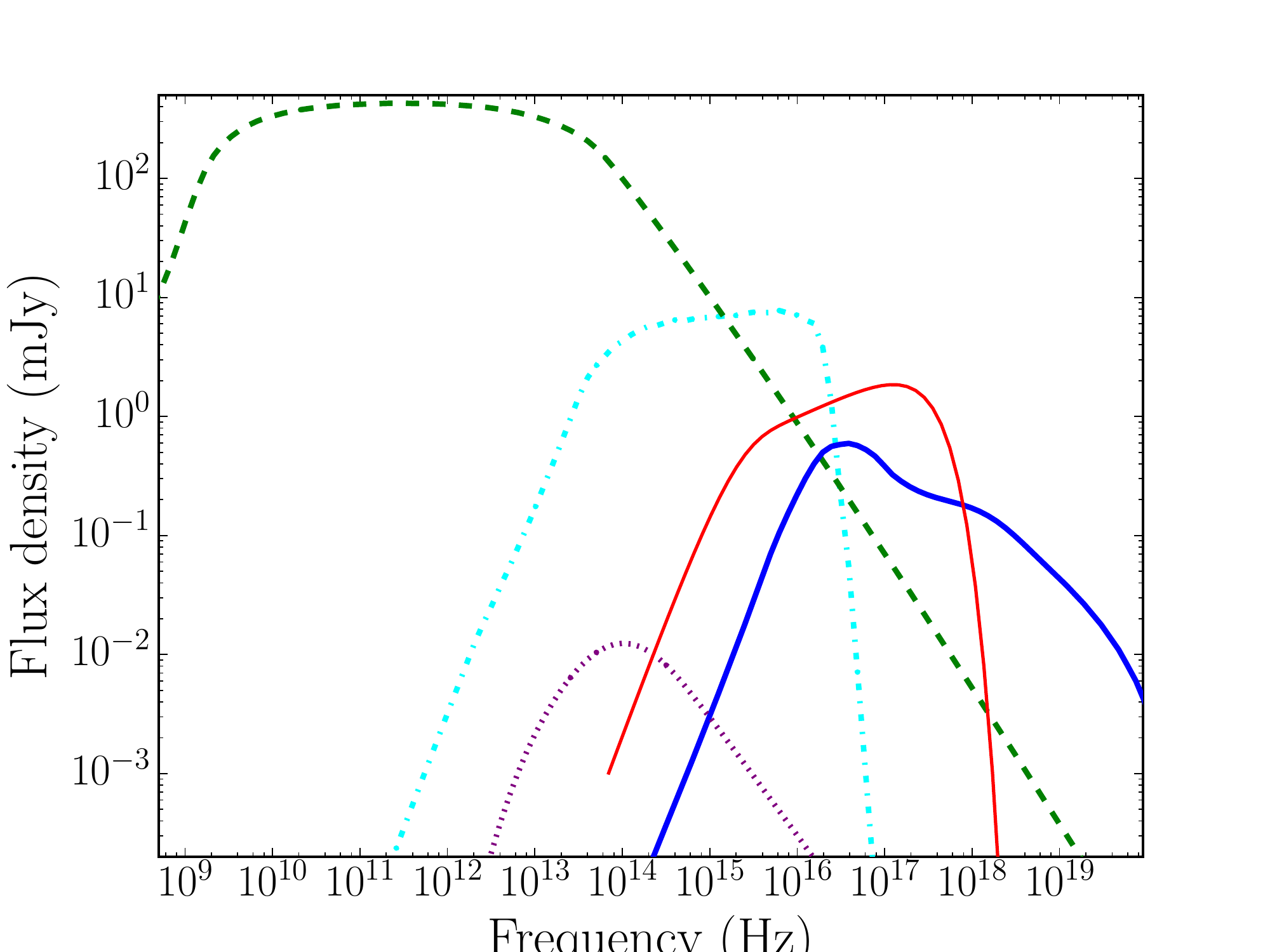}
\caption{Example jet SEDs for different jet powers and pair contents, demonstrating the model degeneracy between jet power, pair content and jet base geometry. For the top left SED, $N_{\rm j} = 4\cdot10^{-2}\,L_{\rm Edd}$, $r_{\rm 0} = 10\,\rm{R_{g}}$, and 20 pairs per proton are injected in the jet. The top right SED uses the same jet power and jet base radius, but one pair per proton. The bottom left SED uses one proton per electron,  $N_{\rm j} = 4\cdot10^{-1}\,L_{\rm Edd}$, and $r_{\rm 0} = 10\,\rm{R_{g}}$. Finally, the bottom right SED uses  $N_{\rm j} = 4\cdot10^{-1}\,L_{\rm Edd}$, and $r_{\rm 0} = 4\,\rm{R_{g}}$. The dashed green and dotted purple lines represent non-thermal synchrotron or inverse-Compton emission; the dotted cyan and continuous thick blue lines represent thermal cyclo-synchrotron or inverse-Compton emission from the jet base; the continuous thin red line the emission from the disk.} 
\label{fig:eta2}
\end{figure*}

The (weaker) correlation between the X-ray variability properties and the particle acceleration location $z_{\rm acc}$ is less intuitive. We also note that due the relatively poor constraints available on the break frequency on September 14 and 16 \citep{Russell14b}, this second result should be interpreted with additional care. However, the existence of such a correlation is not entirely unexpected, as the disk and jet are coupled to each other. From a purely hydrodynamical point of view, one way to explain such a correlation is to assume that propagating fluctuations in the accretion flow, which set the shape of the power spectrum (\citealt{Ingram11}, \citealt{Ingram13}, \citealt{Rapisarda14}, \citealt{Rapisarda16}), also set the details of the internal shocks powering particle acceleration \citep{Malzac13,Peault19}. This could naturally lead to the observed correlation. However, such a scenario does not account for the role that magnetic fields play in determining the dynamics of accretion flows and outflows. From a magneto-hydrodynamics point of view, the location of particle acceleration may be associated with the modified fast point (MFP) of the jet, which can be understood as the location beyond which the outflow ceases to be causally connected with the black hole (e.g. \citealt{Vlahakis04}). In simplified, ideal MHD treatments, the location of the MFP can be uniquely determined when the conditions at the base of the jet are specified, and small changes e.g. in the magnetic field configuration at the jet launching point can change the MFP by orders of magnitude (e.g. \citealt{Ferreira95,Polko10,Polko13,Polko14}). In this sense, a change in the physical conditions at the base of the jet and/or innermost regions of the accretion flow, like the magnetic field configuration, could drive large changes in both the location of the MFP and the properties of the turbulence driving accretion, producing the correlations between timing and spectral parameters we found. Regardless of origin, this new empirical trend provides an interesting benchmark for more complicated numerical studies of dissipation \citep[e.g.][]{Chatterjee19}. A more explicit comparison is currently not possible, as both simulations and semi-analytical GRMHD have yet to fully explore their parameter spaces, due to the computational costs involved. 

Finally, our fits suggest that the jet power in the source is roughly constant and on the order of $\approx 10^{-2}\,L_{\rm Edd}$, and that the SEDs in every epoch can be fitted with a moderate terminal Lorenz factor of $\gamma \approx 3$. These findings are in contrast with \cite{Peault19}, whose model fitted to the same data requires a wide range in both jet powers ($\approx 10^{-3}$ to a few $10^{-1}\,L_{\rm Edd}$) and Lorenz factor (1.05-17). However, we note that in both models jet speed and power are highly degenerate both with each other and with additional parameters (e.g. opening angle, pair content, efficiency of particle acceleration, as shown in Fig.6 in \citealt{Peault19} and discussed in the following section). These degeneracies mean that the best-fitting values of the jet powers should only be considered as order of magnitude estimates. Otherwise, our treatment of the jet is complementary to that of \cite{Peault19}: in their model, the mechanism responsible for particle acceleration is treated in more detail; in ours, we include inverse Compton emission from the jet base and couple it to the full compact jet. Nevertheless, both in our model and theirs, several parameters need to vary between epochs in order to correctly reproduce the data.

\subsection{Pair loading in the jet}
\label{sec:pair_degeneracy}

As mentioned in section \ref{sec:pairs}, in every fit we fixed the initial plasma-$\beta$ parameter  $\beta_{\rm e,0} = 0.02085$,  resulting in a roughly constant pair content of $\approx$ 20 pairs per proton. We chose this target ratio based on a comparison of the energetics of the jet and disk emission. We note that this is an ad-hoc assumption we made purely to reduce the model's degeneracy; however, it has important consequences for the energetic balance of inflowing vs outflowing material.

In the framework of a leptonic model, assuming one cold proton per radiating electron leads to the highest possible jet power for a given observed jet luminosity. Such an assumption leads to reasonable energy budgets in canonical blazars; for example, \cite{Ghisellini14} found that the jet power is on the order of, but slightly higher than, the disk luminosity. A large pair content in the jets in Flat Spectrum Radio Quasars is also disfavoured because of the effect that radiation pressure would have on the outflow dynamics \citep{Ghisellini10}.

In our modelling we found, however, that assuming a low pair content in the jet ($\eta \leq 20$) leads to uncomfortably high energy requirements in the jet. Since the jet emission of this source is not significantly different from other XRBs,  a similar conclusion would also apply to the XRB population as a whole. This discrepancy in required power is illustrated in Fig.\ref{fig:eta2}, which shows the jet+disk SEDs for a constant disk luminosity $L_{\rm disk} = 4\cdot 10^{-2} L_{\rm Edd}$, while varying the jet power and matter content. If the jet carries $\approx 20$ pairs per electron, the jet power required to produce a ``typical'' SED (in terms of radio and X-ray luminosities) is $N_{\rm j} = 4\cdot 10^{-2} L_{\rm Edd}$, as shown in the top left panel; this value is comparable to the disk luminosity, as inferred for blazar jets, and comfortably sub-Eddington as expected for HS XRBs. Instead, if the jet carries one proton per electron, as shown in the bottom right panel, the required jet power to produce a typical BHXB SED is $N_{\rm j} \approx 4\cdot 10^{-1} L_{\rm Edd}$, 20 times higher than the disk luminosity.  

This degeneracy can be understood by writing the jet power for a fixed lepton number density, as a function of pair content $\eta$. By combining eq.\ref{eq:Nj}, \ref{eq:plasma_beta} and \ref{eq:pairs} we get: 
\begin{equation}
N_{\rm j} = 2\gamma_0 \beta_0 c \pi R_0^{2} U_{\rm e,0} \left(1+\frac{\sigma_{\rm 0} m_{\rm p} + \Gamma_{\rm ad} \sigma_{\rm 0} \eta \langle \gamma \rangle m_{\rm e}}{2 \eta \langle \gamma \rangle m_{\rm e}} + \frac{m_{\rm p}}{\eta \langle \gamma \rangle m_{\rm e}} \right).
\end{equation}
The term in brackets is a monotonically decreasing function of $\eta$, implying that very similar SEDs can be produced by reducing $\eta$ and increasing $N_{\rm j}$, regardless of whether it is the protons or leptons that carry the bulk of the kinetic energy.

Jet powers in excess of the accretion power can be achieved through the Blandford-Znajek process \citep{Blandford77}; however, GRMHD simulations find that the efficiency of such a process is on the order of a few at most (e.g. \citealt{Tchekhovskoy11}), far lower than the (conservative) factor $\approx 20$ estimated above. Despite the large uncertainties arising from directly comparing GRMHD simulations to observations, to date this is the best estimate of the efficiency of the Blandford-Znajek process. These energetic considerations imply one of two things: either black hole X-ray binaries are far more efficient at launching jets than blazars are, or the composition of the plasma of the outflows in the two classes of sources is different, with XRBs carrying more pairs than AGN. 

One possible explanation is that the energy densities near the black hole scale inversely with black hole mass, similarly to the peak temperature of an accretion disk, for a fixed Eddington accretion power. Such an increase could lead to an increased efficiency in pair-production processes near the black hole. This scaling can be shown by once again inverting equation \ref{eq:Nj} to solve for the energy density:
\begin{align}
U_{\rm tot} & = \frac{N_{\rm j}}{\pi \gamma_{\rm 0} \beta_{\rm 0} c r_{\rm 0}^{2}}
\nonumber
\\ & = 1.4\times10^{17} \left(\frac{N_{\rm j}}{1.26\times10^{38}\,\rm{erg\,s^{-1}}}\right) \left(\frac{1.5\times10^{5}\,\rm{cm}}{r_{\rm 0}}\right)^{2}.
\end{align}
Taking $N_{\rm j} = 10^{-2}\,\rm{L_{Edd}}$ and $r_{\rm 0} = 10\,\rm{R_g}$, resulting in $U_{\rm tot} = 1.4\times10^{12}$ $\rm{erg}$ $\rm{cm^{-3}}$ for a $10\,M_{\rm \odot}$ black hole, and $U_{\rm tot} = 1.4\times10^{4}$ $\rm{erg}$ $\rm{cm^{-3}}$ for a $10^{9}\,M_{\rm \odot}$ black hole. Therefore, for a given energy budget \textit{in dimensionless Eddington units}, the energy density near a stellar-mass black hole is typically $\sim8$ orders of magnitude larger than that near a large supermassive black hole. Higher energy densities \textit{in physical units} would naturally lead to an increase of pair-production processes near the black hole (e.g. \citealt{Neronov07}, \citealt{Moscibrodzka11}, \citealt{Broderick15}). This in turn could naturally result in XRB jets being more pair-loaded than AGN jets.

\begin{figure}
\includegraphics[width=0.48\textwidth]{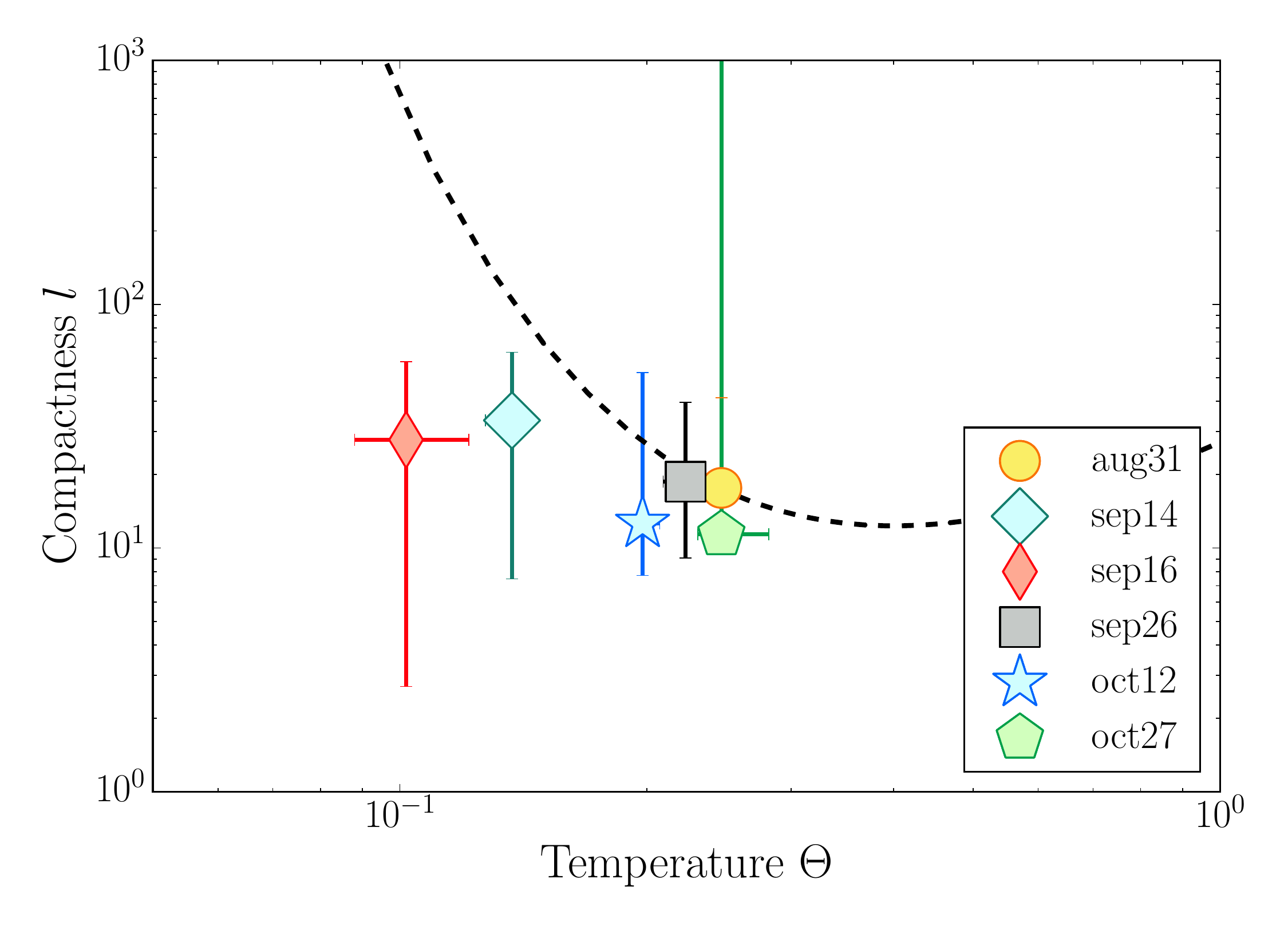}
\caption{Compactness vs dimensionless temperature for the corona, derived from our multi-wavelength fits. The dashed line indicates the pair runaway threshold estimated in \citealt{Svensson84}; each epoch lies close to it, but in the allowed region, implying that some amount of pair production is likely to be occurring in the jet base/corona.}
\label{fig:compactness}
\end{figure}

A useful tool to to quantify the importance of pair production is the compactness-temperature diagram (e.g. \citealt{Fabian15}, and references therein). The dimensionless compactness $l$ is defined as:
\begin{equation}
    l = \frac{L}{R}\frac{\sigma_{\rm t}}{m_{\rm e}c^{3}},
\end{equation}
where $L$ is the luminosity and $R$ is the radius of the emitting region. In the $l-\theta$ (where $\theta$ is the electron temperature in units of 511 keV), regions of high temperature and/or high compactness are forbidden due to pair production becoming a runway process. The exact location of the pair production limit depends on the geometry of the emitting region \citep{Svensson84,Stern95,Svensson96}. For our fits, we computed $L$ by integrating the total model luminosity in the $0.1-300$ keV range, excluding the non-thermal synchrotron emission which originates away from the base of the jet/corona. The resulting $l-\theta$ plot is shown in Fig.\ref{fig:compactness}. All of our fits lie close and to the left of the pair-production region, in excellent agreement with the results of \cite{Fabian15}, suggesting that pair production near the black hole could be regulating the matter content injected in the jet. Older versions of our jet model instead tended to return fits in the runaway region, as discussed by \cite{Malzac09}, further highlighting the importance of treating quasi-relativistic particles in the corona correctly.

\subsection{Implications of a jet-dominated model}

Our proposed scenario for the evolution of the broadband SED of jet-dominated sources makes three predictions:

1) In the harder states, if the cyclo/synchrotron emission of the jet base is brighter than both the companion star and accretion flow, the optical and X-ray emission should be strongly correlated, and both should vary on similar timescales, as both originate in the same region near the black hole. As the source softens and disk irradiation dominates over the cyclo-synchrotron emission of the jet base/corona, the optical should become less variable, and uncorrelated with the hard X-ray power-law.    

2) In the intermediate states, the seed photons for inverse Compton scattering originate in the disk, while in harder states, they originate in the jet base itself. In the first regime, the jet base/corona should react to changes both in the accretion luminosity (which affects the seed photons) and accretion rate, which should impact the mass loading of the jet. In the second case, only mass loading is important, as seed photons are produced locally. These two scenarios would lead to different pivoting behaviour of the X-ray powerlaw, leading in turn to a change in the shape of the reverberation lags \citep{Mastroserio18}.

3) In states closer to quiescence, as the jet power drops, the optical depth should also drop. This can be shown by estimating the dependence of the optical depth in the jet base as a function of jet power and initial jet radius:
\begin{equation}
\tau = n_{\rm e} r_{\rm 0} \sigma_{\rm t}    
\end{equation}
where $n_{\rm e}$ is the electron number density in the jet base, $r_{\rm 0}$ its radius, and $\sigma_{\rm t}$ is the Thomson cross section. We can write $n_{\rm e}$ as a function of the jet power starting from equation \ref{eq:Nj}:
\begin{align}
N_{\rm j} & = 2\pi r_{\rm 0}^{2} \gamma_{\rm 0} \beta_{\rm 0} c \left(U_{\rm e} + U_{\rm b} + U_{\rm p} \right) 
\nonumber
\\ & = 2\pi r_{\rm 0}^{2} \gamma_{\rm 0} \beta_{\rm 0} n_{\rm e} m_{\rm e} \langle \gamma \rangle c^{3} \left(1+\frac{1}{\beta_{\rm e,0}} + \frac{1}{\kappa_{\rm p,0}} \right) 
\label{eq:Nj_2}
\end{align}
where $\langle \gamma \rangle \approx$ a few is the average Lorentz factor of the electrons, $\beta_{\rm e,0} = U_{e}/U_{b}$ and $\kappa_{\rm p,0} = U_{p}/U_{e}$. By making the simplifying assumption that $\langle \gamma \rangle \approx 1$, $\beta_{\rm e,0} = U_{e}/U_{b}$ and $\kappa_{\rm p,0} = U_{p}/U_{e}$ do not change by factors of more than a few as the jet power varies, we find:
\begin{align}
n_{\rm e} = \frac{N_{\rm j}}{2\pi r_{\rm 0}^{2} \gamma_{\rm 0} \beta_{\rm 0} m_{\rm e} \langle \gamma \rangle c^{3} \left(1+\frac{1}{\beta_{\rm e,0}} + \frac{1}{\kappa_{\rm p,0}} \right)} \propto \frac{N_{\rm j}}{r_{0}^{2}}   
\end{align}
and therefore the optical depth scales as:
\begin{equation}
\tau \propto n_{\rm e} r_{\rm 0} \propto \frac{N_{\rm j}}{r_{\rm 0}}.
\end{equation}
The size of the X-ray emitting region in the jet base is highly unlikely to ever become smaller than $\approx 1\,\rm{R_{g}}$, so as the jet power drops at some point the outflow will become increasingly more optically thin, leading to softer Comptonisation spectra. While this can be offset by an increase in temperature, the spectrum of an optically-thin, hot corona is not a power-law, but carries some intrinsic curvature. If a power-law spectrum is to be maintained down to quiescence, therefore, an additional component should become the dominant source of hard X-ray photons - for example, non-thermal synchrotron emission from downstream in the jets. This will also have an impact on the variability of the X-ray continuum, as both the radiative mechanism and emitting region will have changed dramatically from a typical bright hard or hard-intermediate state. Finally, we note that similar conclusions could be drawn in a scenario in which the X-ray powerlaw originates in the RIAF, by replacing $N_{\rm j}$ in equation \ref{eq:Nj_2} with the accretion rate $\dot{M}$. 

To summarise, our jet model predicts that both the characteristics of the X-ray variability should show distinct patterns at different stages of the outburst, tracing the radiative mechanism responsible for the emission in different wavelengths. Furthermore, correlated X-ray and optical variability could be used to discriminate between the latter originating in the accretion flow, companion star or jet base.

\section{Conclusion}
\label{sec:conclusion}

In this paper we have presented an extension of the \texttt{bljet} multi-wavelength, steady-state jet+disk model, and modelled six quasi-simultaneous, multi-wavelength SEDs of the BHXB MAXI J1836$-$194. \texttt{bljet} was initially designed to model the SEDs of blazar jets \citep{Lucchini19a}, starting from the \texttt{agnjet} model \citep{Markoff01,Markoff05} which instead was limited to low-luminosity AGN and BHXBs. Here, we have shown that the \texttt{bljet} formalism can be used to fit the broad-band SED of hard and HIMS states in X-ray binaries.  Furthermore, we combined spectral and timing information in order to provide a more self-consistent picture of the system throughout its outburst. 

We found that in a regime in which the electrons are mildly or non-relativistic at the base of the jet ($T_{\rm e}\approx 100\,\rm{keV}$), this region of the system can produce a power-law coronal continuum by inverse Compton up-scattering either disk-photons (during bright hard and intermediate states), or local cyclo-synchrotron photons (during more faint HSs). We find that the corona/jet launching region is relatively compact, with radii of $\approx 8-25\,\rm{R_g}$; we do not require any additional spectral component to account for the power-law X-ray continuum emission. In our model, if the outflow carries a mild pair content ($\approx$ 20 per proton) we find that the jet power is on the order of, but slightly higher than, the accretion luminosity. A proton-heavy jet carrying no pairs and one proton per electron can in principle reproduce the data, but in this case the jet power required is about 20 times higher than the accretion luminosity; these energetic constraints therefore favour the pair-loaded scenario. 

Our fits indicate that the geometrical properties of the jet are evolving throughout the outburst and are correlated with the shape of the X-ray power spectrum. We find that during softer HIMS states the jet base widens, while the location of particle acceleration in the jet moves downstream, away from the black hole. Vice versa, harder states require more compact jet launching regions and particle acceleration occurring closer to the black hole. We propose that this is caused by a change in the physical conditions at the base of the jet, which cause a variation in both the X-ray variability and jet structure. Such a change in the jet launching conditions could arise, for example, in differences in the collimation environment near the black hole. Recent GRMHD simulations \citep[e.g,][]{Chatterjee19,Liska19} are in broad agreement with this picture. 

Our work shows that combining physically-motivated models, advanced fitting techniques and different observables is a powerful tool to better understand the poorly understood physics of the disk-jet coupling in accreting black holes. In particular, the correlations between timing and spectral properties we have outlined require a time-dependent model to be investigated in depth. This will be the subject of future works.

\section*{Appendix: Radiation calculations}
We include a truncated Shakura-Sunyaev disk \citep{Shakura73}, using the simplified boundary condition \citep{Kubota98} to calculate the normalisation. The disk is parametrised by an inner truncation radius $r_{\rm in}$ and luminosity $L_{\rm d}$; the temperature therefore is:
\begin{equation}
T_{\rm in} = \left(\frac{L_{\rm d}}{2\sigma_{\rm sb}r_{\rm in}^{2}}\right)^{1/4},
\end{equation}
where $\sigma_{\rm sb}$ is the Stefan-Boltzmann constant. We take the disk height at each radius $r$ to be:
\begin{equation}
H = \max (0.1,  L_{\rm d}/L_{\rm Edd})\times r,
\label{eq:diskH}
\end{equation}
where $L_{\rm Edd})$ is the black hole's Eddington luminosity. This results in a thin disk with $H/R = 0.1$ at sub-Eddington luminosities, and a puffed up thicker disk at higher luminosity. We note that for the scope of this work, the disk is always in the thin regime. The only difference in the treatment of the disk is that previous versions of \texttt{agnjet} and \texttt{bljet} considered a fixed height of the disk, while here we fix $H/R$.

Cyclo-synchrotron and inverse Compton emission in \texttt{bljet} are computed following \cite{Blumenthal70}, with some small changes designed to account for emission in the cyclotron regime. In order to account the cyclo-synchrotron emission of both mildly and highly relativistic electrons, we use the phenomenological emissivity formula of \cite{Ghisellini98}:
\begin{equation}
j_{\rm c}^{\prime}(\nu^{\prime},\gamma) = \frac{4\rho^2}{3}
\frac{\sigma_{\rm T}c U_{\rm b}}{\nu_{\rm l}}\frac{2}{1+3\rho^2}
\exp\left[\frac{2(1-\nu^{\prime}/\nu_{\rm l}^{\prime})}{1+3\rho^2}\right]  
\label{eq:cyclo}
\end{equation}
for electrons with $\gamma \leq 2$, and the full synchrotron emissivity
\begin{equation}
j_{\rm s}^{\prime}(\nu^{\prime},\gamma) =  \frac{\sqrt{3}e^{3}B\sin{\theta}}{m_{\rm e}c^{2}} \frac{\nu^{\prime}}{\nu_{\rm s}^{\prime}}\int_{\nu^{\prime}/\nu_{\rm s}^{\prime}}^{\infty} K_{5/3}(y)dy   
\label{eq:syn}
\end{equation}
otherwise. In both equations, $\varrho = p/m_{\rm e}c^{2}$ is the electron momentum in units of $m_{\rm e}c$. Previous versions of \texttt{agnjet} and \texttt{bljet} only included the emissivity in equation \ref{eq:syn}. The emissivity integrated over the particle distribution is:
\begin{equation}
j^{\prime}(\nu^{\prime}) = \int_{\gamma_{\rm min}}^{\gamma_{\rm max}} N(\gamma) j_{\rm c,s}^{\prime}(\nu^{\prime},\gamma) d\gamma,
\end{equation}
where $j_{\rm c,s}^{\prime}(\nu^{\prime},\gamma)$ is the pitch-angle average of equations \ref{eq:cyclo} and \ref{eq:syn}.
The absorption coefficient is:
\begin{equation}
\alpha^{\prime}(\nu^{\prime}) = \int_{\gamma_{\rm min}}^{\gamma_{\rm max}} \frac{N(\gamma)}{\gamma^{2}} \frac{d}{d\gamma}\left( \gamma^{2} j_{\rm c,s}^{\prime}(\nu^{\prime},\gamma)\right) d\gamma  
\label{eq:alfa1}
\end{equation}
as in \cite{Blumenthal70}. The co-moving cyclo-synchrotron luminosity per unit frequency then is:
\begin{equation}
L_{\rm s}^{\prime}(\nu^{\prime}) = \frac{16 \pi H(z)R(z) \nu^{\prime 2}}{c^{2}}\frac{j^{\prime}(\nu^{\prime})}{\alpha^{\prime}(\nu^{\prime})}\left(1-e^{-\tau^{\prime}(\nu^{\prime})} \right),    
\label{eq:syn-comoving}
\end{equation}
where $R(z)$ the radius of the jet at a distance $z$ from the black hole, $H(z)$ is the height of each cylindrical slice of the jet, and
\begin{equation}
 \tau_{\rm s}^{\prime}(\nu^{\prime}) = \frac{\pi}{2 \alpha^{\prime}(\nu^{\prime}) R(z) \delta \sin(\theta)} 
\end{equation}
is the cyclo-synchrotron optical depth, which includes skin depth and viewing angle effects. Finally, the luminosity is boosted in the observer frame with the standard transformations (for a continual jet flow) \citep{Lind85}:
\begin{equation}
L_{\rm s}(\nu) = \delta^{2} L_{\rm c,s}^{\prime}(\nu^{\prime});   \quad \nu = \delta\nu^{\prime}
\label{eq:boost}
\end{equation}

Similarly to cyclo-synchrotron emission, we follow the treatment of \cite{Blumenthal70} to calculate the inverse-Compton emission along the jet, which includes the full Klein-Nishina regime. This is unchanged from previous versions of the model, and is only reported here for completeness. Given an electron with Lorenz factor $\gamma$ and a photon differential density (number of photons per unit volume and per unit of initial photon energy) $N(\epsilon_{\rm 0})$, the scattered photon spectrum in the co-moving frame of the jet is:
\begin{equation}
\frac{dN}{dt d\epsilon_{\rm 1}} = \frac{2\pi r_{\rm e}^{2} m_{\rm e}c^{3} N(\epsilon_{\rm 0}) d\epsilon_{\rm 0}}{\gamma \epsilon_{\rm 0}} F_{\rm IC}^{\prime}(q, \Gamma_{\rm e}) 
\end{equation}
where $\epsilon_{\rm 1}$ is the final photon energy, $\epsilon_{\rm 0}$ is the initial photon energy, $r_{\rm e}$ is the classical radius of the electron, $\gamma$ is the electron's Lorenz factor,  and 
\begin{equation}
F(q, \Gamma_{\rm e}) = 2q\ln{q}+(1+2q)(1-q)+\frac{(\Gamma_{\rm e}q)^{2}(1-q)}{2(1+\Gamma_{\rm e}q)},    
\label{eq:KN}
\end{equation}
where $\Gamma_{\rm e} = 4\epsilon_{\rm 0}\gamma/m_{\rm e}c^{2}$ regulates whether the scattering happens in the Thomson or Klein-Nishina regime ($\Gamma_{\rm e} \ll 1$ and $\Gamma_{\rm e} \gg 1$ respectively), $q = E_{\rm 1} / (\Gamma_{\rm e} (1 - E_{\rm 1}))$ accounts for the photon energy gain from $\epsilon_{\rm 0}$ to $\epsilon_{\rm 1}$, and $E_{\rm 1} = \epsilon_{\rm 1}/\gamma m_{\rm e}c^{2}$ is the final photon energy in units of the initial electron energy. The total spectrum is found by integrating over the particle and seed photon distributions:
\begin{equation}
\frac{dN_{\rm tot}}{dt d\epsilon_{\rm 1}} = \int \int N(\gamma) \frac{dN}{dt d\epsilon_{\rm 1}} d\gamma d\epsilon_{\rm 0},   
\label{eq:IC}
\end{equation}
which has units of total number of scatterings, per unit of outgoing photon energy, volume and time. Finally, the co-moving luminosity is obtained by multiplying equation \ref{eq:IC} by the emitting volume times $h \epsilon_{\rm 1}$, where $h$ is the Planck constant, in order to obtain a specific luminosity ($\rm erg\,s^{-1}\,Hz^{-1}$). The observed spectrum is then calculated with the same Doppler boost as the cyclo-synchrotron one (eq. \ref{eq:boost}).

For inverse-Compton scattering, both disk and cyclo-synchrotron photons produced within the jet are considered. The seed cyclo-synchrotron photon distribution is calculated by converting equation \ref{eq:syn-comoving} into units of number of photons per unit volume and energy:
\begin{equation}
n(\epsilon_{\rm 0}) = \frac{L_{\rm s}^{\prime}(\nu^{\prime})}{ch^{2}\nu^{\prime}\pi R(z)^{2}}.
\end{equation}
The disk seed photon distribution is calculated by integrating the disk temperature profile along the disk radial profile:
\begin{equation}
n_{\rm disk}(\nu, \delta) = \int_{\alpha_{\rm min}}^{\alpha_{\rm max}} \frac{4\pi \nu^{2}(\delta)}{hc^{3}\left(e^{h\nu(\delta) / k_{\rm b}T(\alpha, \delta)} -1\right)} d\alpha,
\end{equation}
where the extremes of the integral are $\alpha_{\rm min} = \arctan(r_{\rm in}/z)$, $\alpha_{\rm max} = \arctan(R_{\rm out}/(z-HR_{\rm out}/2))$ if $z<HR_{\rm out}/2)$ and $\alpha_{\rm max} = \pi/2 \arctan(R_{\rm out}/(z-HR_{\rm out}/2))$ otherwise. In this way, $\alpha$ includes the entire disk viewing angle, from innermost to outermost regions. $r_{\rm in}$ and $R_{\rm out}$ are the inner and outer disk radii, $H$ is the disk aspect ratio defined in equation \ref{eq:diskH}, $z$ is the distance of the jet segment from the black hole, $\nu(\delta)$ is the seed photon frequency in the co-moving frame of the jet, and $T(\alpha, \delta)$ is the observed disk temperature in the co-moving frame of the jet, assuming a standard Shakura-Sunyaev disk temperature profile. 

To calculate multiple scatterings, the IC/SSC spectrum is then passed as the input spectrum back in eq.\ref{eq:IC}:
\begin{align}
\frac{dN_{\rm tot}}{dtd\epsilon_{\rm1}} = \sum_{i} \int \int N(\gamma) \frac{N_i}{dtd\epsilon_{\rm 1}} d\gamma d\epsilon_{\rm 1}
\end{align}
to calculate the total spectrum which is made by the sum of i-scatterings.
We fix the maximum number of iterations/scattering to 15 to maintain a short run time per fit and, at the same time, confidently include all the scattering orders that could be relevant even in the regime of large optical depths. 

\section*{Acknowledgements}
We thank the anonymous referee for useful comments that greatly improved the quality of the manuscript. M. L. and S. M. are thankful for support from an NWO (Netherlands Organisation for Scientific Research) VICI award, grant Nr. 639.043.513. F. V. acknowledges support from STFC under grant ST/R000638/1. TDR acknowledges support from a NWO Veni Fellowship, grant number 639.041.646. This research has made use of ISIS functions (ISISscripts) provided by ECAP/Remeis observatory and MIT (http://www.sternwarte.uni-erlangen.de/isis/). 

\section*{Data Availability}
All data in this paper is publicly available. The radio data was published and tabulated in \cite{Russell14a,Russell14b,Russell15}. The Infrared, optical and UV data was published in \cite{Russell13}. The X-ray data is publicly available from HEASARC (\url{https://heasarc.gsfc.nasa.gov/}). A reproduction package is available at DOI: 10.5281/zenodo.4384809.

\bibliographystyle{mnras}
\bibliography{references}

\end{document}